\documentclass[prb,floatfix,twocolumn,showpacs,amsmath,amssymb]{revtex4}
\usepackage{graphicx}
\usepackage{epstopdf}
\usepackage{epsfig}
\usepackage{dcolumn}
\usepackage{bm}
\usepackage{color}

\newcommand{\dagga}{{\phantom{\dagger}}}

\begin{document}

\title{Variational wave functions for the $S=1/2$ Heisenberg model on the anisotropic triangular lattice: 
Spin liquids and spiral orders}
\author{Elaheh Ghorbani,$^{1,2}$ Luca F. Tocchio,$^1$ and Federico Becca$^1$}
\affiliation{$^1$Democritos National Simulation Center, Istituto Officina dei Materiali del CNR, 
and SISSA-International School for Advanced Studies, Via Bonomea 265, I-34136 Trieste, Italy \\
$^2$Department of Physics, Isfahan University of Technology, Isfahan 84156-83111, Iran}

\date{\today} 

\begin{abstract}
By using variational wave functions and quantum Monte Carlo techniques, we investigate the complete phase diagram of 
the Heisenberg model on the anisotropic triangular lattice, where two out of three bonds have super-exchange couplings
$J$ and the third one has instead $J^\prime$. This model interpolates between the square lattice and the isotropic 
triangular one, for $J^\prime/J \le 1$, and between the isotropic triangular lattice and a set of decoupled chains, 
for $J/J^\prime \le 1$. We consider all the fully-symmetric spin liquids that can be constructed with the fermionic 
projective-symmetry group classification [Y. Zhou and X.-G. Wen, arXiv:cond-mat/0210662] and we compare them with the spiral 
magnetic orders that can be accommodated on finite clusters. Our results show that, for $J^\prime/J \le 1$, the phase 
diagram is dominated by magnetic orderings, even though a spin-liquid state may be possible in a small parameter 
window, i.e., $0.7 \lesssim J^\prime/J \lesssim 0.8$. In contrast, for $J/J^\prime \le 1$, a large spin-liquid region 
appears close to the limit of decoupled chains, i.e., for $J/J^\prime \lesssim 0.6$, while magnetically ordered phases 
with spiral order are stabilized close to the isotropic point.
\end{abstract}


\maketitle

\section{Introduction}\label{sec:intro}

The field of frustrated magnetism represents an active research topic in condensed matter physics,~\cite{lacroix2011} 
due to the possibility that unconventional phases may be stabilized, with topological properties and fractionalized 
excitations (i.e., carrying fractions of the ``elementary quantum numbers'' or obeying fractional or anyonic statistics).
Fascinating examples are given by the so-called spin liquids, which are obtained whenever competing magnetic interactions 
are strong enough to prevent any possible magnetic ordering down to zero temperature. Among various materials that show
promising low-temperature behaviors, the family of organic charge-transfer salts $\kappa$-(ET)$_2$X represents a very 
important candidate for hosting spin-liquid properties. In these materials, strongly dimerized organic molecules are 
arranged in stacked two-dimensional anisotropic triangular layers, where, up to a good level of approximation, two out of 
three hoppings are equal ($t$), while the third one is different ($t^\prime$). Several magnetic and superconducting phases 
may be observed, by varying pressure and temperature, as well as the nature of the anion X.~\cite{kanoda2011,powell2011} 
The most interesting compound of the family is given by $\kappa$-(ET)$_2$Cu$_2$(CN)$_3$, where no signal of magnetic order 
has been detected down to very low temperatures, thus indicating that a non-magnetic Mott insulator may be eventually 
realized.~\cite{shimizu2003,manna2010} Since this compound is moderately correlated, the estimate of the frustrating ratio 
has been performed by several independent groups within density-functional approaches, which estimated the values of the 
two different hoppings $t$ and $t^\prime$. These calculations lead to a degree of anisotropy that is between the square 
lattice and the isotropic triangular one, e.g., $t^\prime/t \simeq 0.85$.~\cite{kandpal2009,nakamura2009,scriven2012}
However, this result has been recently questioned by a calculation that instead suggests a more isotropic triangular 
structure, e.g., $t^\prime/t \simeq 1$.~\cite{koretsune2014} In any case, the $\kappa$-(ET)$_2$X family is a fertile 
field for the search of spin-liquid compounds. Indeed, besides $\kappa$-(ET)$_2$Cu$_2$(CN)$_3$, compounds have been recently
discovered, showing interesting low-temperature properties; among them, we would like to mention $\kappa$-(ET)$_2$B(CN)$_4$,
which appears to be strongly correlated and with a marked one-dimensional character.~\cite{yoshida2015} Finally, also 
another family of salts based on the organic molecule Pd(dmit)$_2$ has been shown to possess rich phase 
diagrams.~\cite{kanoda2011} Even if a fully anisotropic triangular lattice is more suitable to properly describe 
these materials,~\cite{scriven2012,jacko2013} a simpler modelization in terms of a $t{-}t^\prime$ anisotropic triangular 
lattice has been also considered in density-functional theory.~\cite{scriven2012} The estimated hopping parameters fall in 
the window $0.75\lesssim t^\prime/t\lesssim 0.9$ for the two compounds with a non-magnetic ground state, namely 
Me$_3$EtSb[Pd(dmit)$_2$]$_2$~\cite{itou2008} and Me$_3$EtP[Pd(dmit)$_2$]$_2$.~\cite{tamura2006}

Besides these organic materials, the anisotropic triangular lattice is also appropriate to describe two isostructural 
and isoelectronic compounds, Cs$_2$CuBr$_4$ and Cs$_2$CuCl$_4$, where magnetic Copper atoms lie on weakly coupled triangular
lattices. While Cs$_2$CuCl$_4$ shows spin-liquid behavior over a broad temperature range, with spin excitations that appear 
to be gapless,~\cite{coldea2001,vachon2011} the Cs$_2$CuBr$_4$ compound exhibits a magnetic ground state with spiral order 
in zero magnetic field.~\cite{ono2004} Both these materials are much more strongly correlated than organic salts and their
physical properties can be captured by frustrated Heisenberg models (possibly decorated by small perturbations, such as 
the Dzyaloshinskii-Moriya interaction). Most importantly, these materials have $J<J^\prime$ and, therefore, their 
structure can be seen as chains (defined along $J^\prime$) coupled together through a zigzag coupling ($J$).
The distinct physical behaviors are generally attributed to the different degree of frustration, i.e., the ratio between 
inter-chain and intra-chain super-exchange couplings in the underlying anisotropic triangular lattice. For Cs$_2$CuCl$_4$, 
a direct comparison between neutron scattering experiments and theoretical calculations,~\cite{coldea2001} the temperature 
dependence of the magnetic susceptibility,~\cite{zheng2005} as well as recent estimates based on spin-resonance 
spectroscopy experiments,~\cite{zvyagin2014} suggest that the ratio between intra-chain $J^\prime$ and inter-chain $J$ 
magnetic couplings is $J/J^\prime \simeq 0.33$. A small inter-layer coupling $J_{\perp}$ of the order of $10^{-2}J^\prime$ 
is responsible for the appearance of a three-dimensional magnetic order below $T_N=0.62$K. Instead, the situation is more 
controversial for the Cs$_2$CuBr$_4$ compound, where a comparison of the experimental results with the theoretical 
calculations of Ref.~\onlinecite{weihong1999} suggests a frustrating ratio of $J/J^\prime \simeq 0.75$,~\cite{ono2005} in 
agreement with density-functional theory calculations,~\cite{foyevtsova2011} while a more recent determination of the 
super-exchange couplings via spin-resonance spectroscopy indicates more one-dimensional features, i.e., 
$J/J^\prime \simeq 0.4$.~\cite{zvyagin2014} 

In this paper, we study the frustrated $S=1/2$ Heisenberg model on the anisotropic triangular lattice. Despite its 
simplicity, the ground state of this model is still controversial, with different methods giving more emphasis either to 
spin liquids or to magnetic phases. On the one hand, spiral orders with non-trivial periodicities appear at the classical 
level and may survive to quantum fluctuations; previous calculations have considered this issue within various 
approximations, e.g., by using series expansion,~\cite{weihong1999} coupled-cluster approaches,~\cite{bishop2009} and the 
Gutzwiller approximation.~\cite{powell2007} Recently, the density-matrix renormalization group (DMRG) has been used to study 
magnetic correlations and the associated finite-size effects, obtaining an incommensurate behavior over a wide range of the 
phase diagram.~\cite{weichselbaum2011} On the other hand, the presence of competing interactions in the anisotropic 
triangular lattice attracted also a large interest in the search of spin-liquid phases. While the isotropic point is well 
established as a magnet with a three-sublattice periodicity (the so-called $120^{\circ}$ 
order),~\cite{bernu1994,capriotti1999,white2007} the presence of anisotropies may favor a non-magnetic ground state with 
respect to generic spiral states. However, on the anisotropy regime interpolating between a square lattice and the isotropic
triangular one, the quest for spin-liquid phases has been very limited and includes only calculations based upon 
spin-wave theories~\cite{merino1999,trumper1999,hauke2011,holt2014} or on Schwinger bosons.~\cite{manuel1999} 
A dimer-ordered state has been also proposed for $0.7\lesssim J^\prime/J\lesssim 0.9$ by a series-expansion 
approach.~\cite{weihong1999} On the contrary, when the anisotropic triangular lattice interpolates between the triangular 
lattice and a set of decoupled chains, the existence of an essentially one-dimensional spin-liquid phase for 
$J/J^\prime \ll 1$ has been investigated by using variational Monte Carlo (VMC) based on Gutzwiller projected 
states,~\cite{yunoki2006,hayashi2007,heidarian2009} exact diagonalization,~\cite{weng2006} and the functional renormalization
group,~\cite{reuther2011} the last study also including the presence of incommensurate magnetism, close to the isotropic 
point. The strong one-dimensional nature of this spin-liquid phase has been also investigated by a mean-field study based 
on Majorana fermions.~\cite{herfurth2013} An alternative scenario has been suggested in the limit of quasi-one dimensional 
lattices, where magnetic order with a collinear pattern could be stabilized;~\cite{starykh2007} this claim has been supported
by using coupled-cluster methods~\cite{bishop2009} and DMRG calculations on a three-leg ladder.~\cite{chen2013} Some evidence
that, for $J/J^\prime \ll 1$, collinear antiferromagnetism is favored over generic spiral states with coplanar order has 
been also reported by an exact diagonalization study with twisted boundary conditions.~\cite{thesberg2014} Finally, in 
addition to the one-dimensional (gapless) spin liquid, VMC calculations also suggested the possibility that a (gapped) spin 
liquid exists close to the isotropic point.~\cite{yunoki2006,heidarian2009} However, in this regime, a direct comparison 
with magnetically ordered states with spiral order was not fully considered.

The aim of this work is to perform a direct comparison of different spin liquids and magnetic states with spiral order, 
that are treated on the same ground within the VMC approach, thus going beyond the previous limitations. We draw the 
complete phase diagram of the Heisenberg model on the anisotropic triangular lattice, for both $J^\prime/J \le 1$ and 
$J/J^\prime \le 1$. With respect to previous VMC works that considered complicated parametrizations of the wave functions 
with several variational parameters,~\cite{yunoki2006,heidarian2009} here we construct more transparent wave functions for 
both spiral magnetic order and non-magnetic states. In particular, spiral phases are considered for the first time; 
moreover, we analyze the spin liquids that can be constructed by using the fermionic projective-symmetry group 
classification.~\cite{zhou2003} The main result of our calculations is that magnetic states with spiral order have 
competing energies with respect to magnetically disordered states in the whole phase diagram. For $J^\prime/J \le 1$, 
magnetic states always have a lower energy compared to spin liquids, which are competitive only in a small window 
$0.7 \lesssim J^\prime/J \lesssim 0.8$. For $J/J^\prime \le 1$, we confirm that, close to the limit of decoupled chains, 
spin-liquid wave functions have better energies with respect to magnetic states (including collinear ones), indicating the 
presence of a quasi-one-dimensional magnetically disordered phase for $J/J^\prime \lesssim 0.6$. In contrast, close to the 
isotropic point, spiral states have lower energies than spin-liquid ones. 

The paper is organized as follows: in Sec.~\ref{sec:model}, we introduce the Heisenberg model and the variational wave 
functions that are constructed for magnetic and spin-liquid states; in Sec.~\ref{sec:results}, we present the results of 
our numerical calculations; finally, in Sec.~\ref{sec:conc}, we draw our conclusions.   

\section{Model and method}\label{sec:model}

\begin{figure}
\includegraphics[width=0.65\columnwidth]{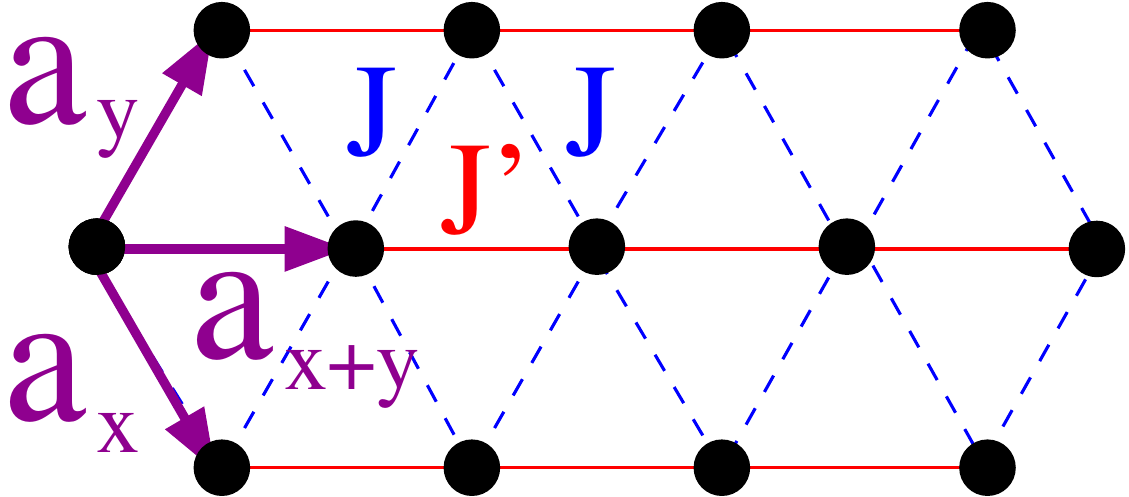}
\caption{\label{fig:lattice}
(Color online) Illustration of the anisotropic triangular lattice, where red solid and blue dashed lines indicate 
antiferromagnetic couplings $J^\prime$ and $J$, respectively.} 
\end{figure}

We consider the spin-1/2 antiferromagnetic Heisenberg model on the anisotropic triangular lattice, as described by:
\begin{equation}
{\cal H}=\sum_{i,j}J_{ij}{\bf S}_i{\bf S}_j,
\end{equation}
where ${\bf S}_i=(S_i^x,S_i^y,S_i^z)$ is the spin-1/2 operator at the site $i$ and $J_{ij}$ is the antiferromagnetic 
coupling, including an intra-chain $J^\prime$ along ${\bf a}_{x+y}=(1,0)$, and an inter-chain $J$ along 
${\bf a}_{x}=(1/2,-\sqrt{3}/2)$ and ${\bf a}_{y}=(1/2,\sqrt{3}/2)$, see Fig~\ref{fig:lattice} and Ref.~\onlinecite{zhou2003}. 
The coordinates of the lattice sites are given by ${\bf R}_i=i_x {\bf a}_{x} + i_y {\bf a}_{y}$. In the following, we consider 
clusters with periodic boundary conditions defined by the vectors ${\bf T}_{x}=l{\bf a}_{x}$ and ${\bf T}_{y}=l{\bf a}_{y}$, 
which contain $L=l^2$ sites. We will take $J$ as the unit of energies for the region of the phase diagram
with $J^\prime/J <1$ and $J^\prime$ for the region with $J/J^\prime<1$.

Our numerical results are based on the definition of correlated variational wave functions that approximate the exact
ground-state properties. In particular, they are given by:
\begin{equation}\label{eq:psi}
|\Psi\rangle = {\cal J}_s {\cal P}_G|\Phi_0\rangle,
\end{equation}
where ${\cal P}_G=\prod_{i}(1-n_{i,\uparrow}n_{i,\downarrow})$ is the usual Gutzwiller projection operator onto the subspace 
of singly occupied sites, with the total number of electrons $N$ equal to the number of sites $L$. Moreover, ${\cal J}_s$ 
is a spin-spin Jastrow term:
\begin{equation}
{\cal J}_s = \exp \Big[ \frac{1}{2} \sum_{i,j} u_{ij} S_i^z S_j^z \Big], 
\label{eq:jastrowspin} 
\end{equation}
where $u_{ij}$ are pseudo-potentials that can be optimized for every independent distance $|{\bf R}_i-{\bf R}_j|$ in order 
to minimize the variational energy. Finally, $|\Phi_0\rangle$ is the ground state of a non-interacting fermionic Hamiltonian,
which can describe either magnetic or non-magnetic states. All the calculations are performed in the subspace with no net 
spin polarization, i.e. $\sum_i S^z_i=0$. Given the presence of the Gutzwiller projector and the Jastrow factor, a Monte 
Carlo sampling is needed to compute any expectation value over these correlated variational states.

Let us now describe the two families of variational states that are used to draw the phase diagram. The first one is given
by magnetic states. In this case, $|\Phi_0\rangle$ is the ground state of a non-interacting fermionic Hamiltonian that 
contains both a band contribution and a magnetic term:
\begin{equation}
{\cal H}_{\rm AF}={\cal H}_{\rm kin}+{\cal H}_{\rm mag}.
\end{equation}
The magnetic term is of the form 
\begin{eqnarray}
{\cal H}_{\rm mag } &=& 2 h \sum_j \left[ \cos({\bf Q}\cdot{\bf R}_j) S_j^x+\sin({\bf Q}\cdot {\bf R}_j)S_j^y \right] \nonumber \\
&=& h \sum_j \left[ e^{-i {\bf Q}\cdot{\bf R}_j} c^{\dagger}_{j,\uparrow} c^{\dagga}_{j,\downarrow} +
                     e^{i {\bf Q}\cdot{\bf R}_j} c^{\dagger}_{j,\downarrow} c^{\dagga}_{j,\uparrow} \right],
\label{eq:HAF}
\end{eqnarray}
where the pitch vector ${\bf Q}$ determines the magnetic ordering in the $x{-}y$ plane and $h$ is a variational parameter 
to be optimized. Here, the fermionic operator $c^{\dagger}_{j,\sigma}$ ($c^{\dagga}_{j,\sigma}$) creates (destroys) one
electron with spin $\sigma$ on the site $j$.

\begin{figure}
\includegraphics[width=\columnwidth]{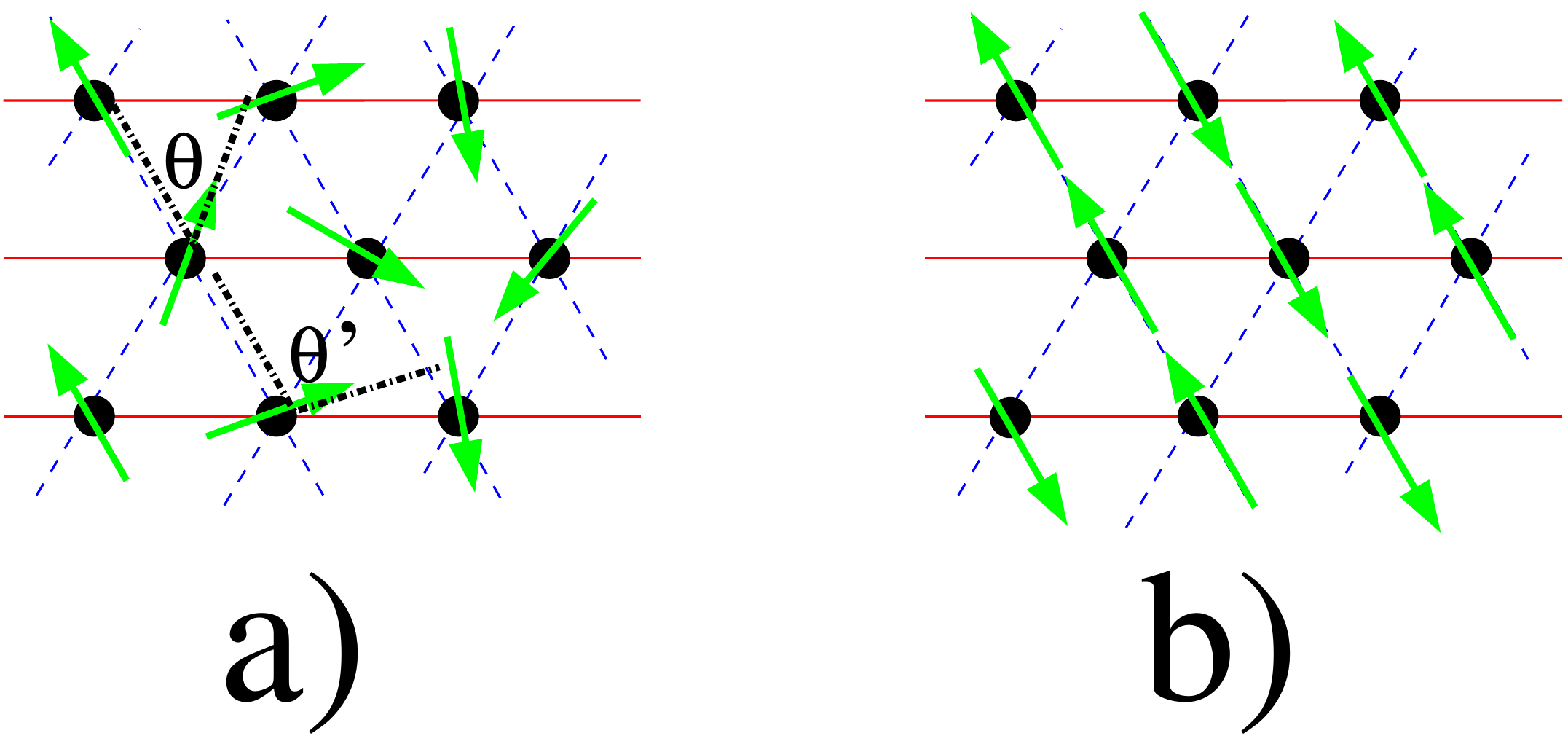}
\caption{\label{fig:orders}
(Color online) Left panel: spin pattern for a spiral state with $\theta^\prime=2\theta$. Right panel: spin pattern for the 
collinear state proposed to describe the limit $J \to 0$.~\cite{starykh2007}} 
\end{figure}

In real space, the magnetic order can be described by two angles $\theta$ and $\theta^\prime$, defining the relative 
orientation of two neighboring spins along ${\bf a}_y$ and ${\bf a}_{x+y}$, respectively. According to previous 
calculations,~\cite{weihong1999,bishop2009} the optimal magnetic solution displays a spiral order, which may be parametrized 
through a single angle $\theta\in[\pi/2,\pi]$, with $\theta^\prime=2\theta$, see Fig~\ref{fig:orders} left panel. 
For example, a pitch angle of $\theta=\pi$ corresponds to N\'eel order, appropriate for the limit $J^\prime \to 0$, while 
$\theta=2\pi/3$ corresponds to the $120^{\circ}$ order, suitable for $J=J^\prime$. In addition to these two magnetic 
orderings, we consider in our calculations few intermediate spiral orders, as allowed by the size of the clusters on which 
the simulations are performed. Indeed, given an $l\times l$ cluster with periodic boundary conditions, the allowed $\theta$'s 
are given by $\theta=2\pi n/l$, with $n$ being an integer. Besides this class of spiral states, we also consider states with 
collinear order in the limit $J \to 0$, i.e., states with $\theta^\prime=\pi$ and $\theta=0$ or $\pi$, see Fig~\ref{fig:orders}
right panel, as suggested in Ref.~\onlinecite{starykh2007}.

The kinetic part of the Hamiltonian is given by:
\begin{equation}\label{eq:kin}
{\cal H}_{\rm{kin}}=\sum_{i,j,\sigma} \chi_{ij} c^{\dagger}_{i,\sigma}c^{\dagga}_{j,\sigma} + {\rm h.c.},
\end{equation}
where $\chi_{ij}$ are hopping parameters that connect nearest-neighbor sites. We considered two possible {\it Ans\"atze} that
describe the case with vanishing magnetic fluxes:
\begin{equation}\label{eq:hopping_1x1}
\chi_{ij}=\left\{ 
\begin{array}{ll}
  \chi        & {\rm for} \;\;\; {\bf R}_j={\bf R}_i+{\bf a}_{x} \\
  \chi        & {\rm for} \;\;\; {\bf R}_j={\bf R}_i+{\bf a}_{y} \\
  \chi^\prime & {\rm for} \;\;\; {\bf R}_j={\bf R}_i+{\bf a}_{x+y} \\
  0           & \textrm{otherwise}
\end{array} 
\right.,
\end{equation}
and the case with $\pi$-fluxes threading up triangles (and $0$-flux threading down triangles):
\begin{equation}\label{eq:hopping_2x1}
\chi_{ij}=\left\{ 
\begin{array}{ll}
  \chi                  & {\rm for} \;\;\; {\bf R}_j={\bf R}_i+{\bf a}_{x} \\
  -(-1)^{i_x}\chi       & {\rm for} \;\;\; {\bf R}_j={\bf R}_i+{\bf a}_{y} \\
  (-1)^{i_x}\chi^\prime & {\rm for} \;\;\; {\bf R}_j={\bf R}_i+{\bf a}_{x+y} \\
   0                    & \textrm{otherwise} 
\end{array} 
\right.,
\end{equation}
where (in both cases) $\chi^\prime$ is a variational parameter to be optimized, while $\chi=1$ in order to set the energy 
scale. Remarkably, the trivial case with no fluxes of Eq.~(\ref{eq:hopping_1x1}) gives the best {\it Ansatz} only for 
$J/J^\prime \lesssim 0.5$, while for $J/J^\prime \gtrsim 0.5$ the best magnetic state is obtained with ${\cal H}_{\rm kin}$ 
having non-trivial $\pi$-fluxes piercing the lattice. The fact of having a non-trivial pattern of magnetic fluxes in the 
kinetic part of the non-interacting Hamiltonian could be surprising. However, on the square lattice, we have shown that the 
best variational wave function can be constructed by considering both N\'eel order and superconducting pairing with 
$d_{x^2-y^2}$ symmetry in the non-interacting Hamiltonian,~\cite{lugas2006,spanu2008} which is equivalent to having non-trivial 
magnetic fluxes. Also in the triangular lattice, previous calculations~\cite{yunoki2006,heidarian2009} have shown that the 
best state is obtained with a $2 \times 1$ unit cell in the non-interacting Hamiltonian, implying non-trivial fluxes. In fact, 
classical magnetic order alone does not reproduce the correct signs of the ground state and the $2 \times 1$ unit cell
strongly improves them.~\cite{capriotti1999,yunoki2006} We want to stress the fact that, in all cases considered here, the 
non-interacting Hamiltonian has a finite gap, due to the presence of antiferromagnetic order; this fact implies that the 
variational state describes a conventional magnetic state. Finally, we remark that a spin Jastrow factor, coupling the 
$z$-component of the spins when magnetic order is defined in the $x{-}y$ plane, is fundamental to reproduce the spin-wave 
fluctuations above the magnetic mean-field state.~\cite{franjic1997,becca2000}

The spin-liquid wave functions are constructed from the classification obtained in Ref.~\onlinecite{zhou2003}. The starting
point is the most general form of an $SU(2)$ invariant mean-field Hamiltonian, of which $|\Phi_0\rangle$ would be the ground 
state:
\begin{equation}\begin{split}\label{eq:H_BCS}
{\cal H}_{\rm SL} =& -\sum_{\langle i,j \rangle} \left( c^{\dagger}_{i,\uparrow} c^{\dagga}_{i,\downarrow}\right) U_{ij}
\left(\begin{array}{c} c^{\dagga}_{j,\uparrow}\\ c^{\dagger}_{j,\downarrow}\end{array}\right) +\rm{h.c.} \\
                   & + \sum_i \sum_{l=1,2,3} \left( c^{\dagger}_{i,\uparrow} c^{\dagga}_{i,\downarrow}\right) a_0^l \tau^l 
\left(\begin{array}{c} c^{\dagga}_{i,\uparrow}\\ c^{\dagger}_{i,\downarrow}\end{array}\right),
\end{split}
\end{equation}
where $U_{ij}$ are written in terms of the Pauli matrices $\tau^l$ ($l=1$, $2$, $3$) and the identity matrix $\mathbb{I}$:
\begin{equation}\label{eq:Umatrix}
U= i t_I \mathbb{I} + t_R \tau^3 + \Delta_I \tau^2 + \Delta_R \tau^1,
\end{equation}
where the bond index $(ij)$ has been dropped for simplicity. Here, $t_R$, $t_I$, $\Delta_R$, and $\Delta_I$ are variational 
parameters to be optimized, as well as the $a_0^l$ of Eq.~(\ref{eq:H_BCS}). We remark that the term proportional to $\tau^3$ 
($\mathbb{I}$) in Eq.~(\ref{eq:Umatrix}) represents kinetic energy with real (imaginary) hopping, while the term proportional 
to $\tau^1$ ($\tau^2$) represents real (imaginary) BCS pairing.

Starting from Eq.~(\ref{eq:H_BCS}), the authors of Ref.~\onlinecite{zhou2003} classified spin-liquid states by using the
projective-symmetry group analysis. Once limiting to solutions having finite couplings $U_{ij}$ along ${\bf a}_{x}$, 
${\bf a}_{y}$, and ${\bf a}_{x+y}$, it was found that the anisotropic triangular lattice can accommodate seven independent
$Z_2$ spin liquids and three $U(1)$ spin liquids. They are labelled with $A$, if translationally invariant, and with $B$, if 
defined on a $2\times 1$ unit cell (in the presence of the Gutzwiller projector they are always totally symmetric). We implemented 
all these states in our Monte Carlo calculations~\cite{ph} and found that the following three {\it Ans\"atze} are relevant 
in some region of the phase diagram: 
\begin{itemize}
\item The $U(1)$ Dirac spin liquid with a $2\times 1$ unit cell and denoted by $U1B\tau^1 \tau^0_{-} \tau^1_{+}$ in 
Ref.~\onlinecite{zhou2003} (relevant for $J^\prime \simeq J$)
\begin{equation}\label{eq:SL1}
U_{ij}=\left\{
\begin{array}{ll}
  \chi\tau^3              & {\rm for} \;\;\; {\bf R}_j={\bf R}_i+{\bf a}_{x} \\
  -(-1)^{i_x}\chi\tau^3   & {\rm for} \;\;\; {\bf R}_j={\bf R}_i+{\bf a}_{y} \\
  (-1)^{i_x}\lambda\tau^3 & {\rm for} \;\;\; {\bf R}_j={\bf R}_i+{\bf a}_{x+y} \\
  0                       & \textrm{otherwise}
\end{array}
\right.,
\end{equation}
together with $a_0^{1,2,3}=0$. This state has gapless excitations with four Dirac points.
\item The $Z_2Ad$ spin liquid with a $1\times 1$ unit cell and denoted by $Z2A\tau^1 \tau^1_{+} \tau^3_{+}$ in
Ref.~\onlinecite{zhou2003} (relevant for $J^\prime/J<1$)
\begin{equation}\label{eq:SL2}
U_{ij}=\left\{
\begin{array}{ll}
  \chi\tau^1+\eta\tau^2   & {\rm for} \;\;\; {\bf R}_j={\bf R}_i+{\bf a}_{x} \\
  \chi\tau^1-\eta\tau^2   & {\rm for} \;\;\; {\bf R}_j={\bf R}_i+{\bf a}_{y} \\
  \lambda\tau^1           & {\rm for} \;\;\; {\bf R}_j={\bf R}_i+{\bf a}_{x+y} \\
  0                       & \textrm{otherwise}
\end{array}
\right.,
\end{equation}
together with $a_0^1=a_1$ and $a_0^{2,3}=0$. For $\lambda=0$ this {\it Ansatz} coincides with the well-known $d$-wave state 
that has been widely used to study the Heisenberg model on the square lattice,~\cite{gros1988,zhang1988,powell2007} which 
is indeed obtained with $J^\prime=0$. For the optimal variational parameters in all the range $J^\prime/J<1$ this state is 
gapless with four Dirac points.
\item The $Z_2As$ spin liquid with a $1 \times 1$ unit cell and denoted by $Z2A\tau^0 \tau^0_{+} \tau^3_{+}$ 
in Ref.~\onlinecite{zhou2003} (relevant for $J/J^\prime<1$)
\begin{equation}\label{eq:SL2_bis}
U_{ij}=\left\{
\begin{array}{ll}
  \chi\tau^1+\eta\tau^2   & {\rm for} \;\;\; {\bf R}_j={\bf R}_i+{\bf a}_{x} \\
  \chi\tau^1+\eta\tau^2   & {\rm for} \;\;\; {\bf R}_j={\bf R}_i+{\bf a}_{y} \\
  \lambda\tau^1           & {\rm for} \;\;\; {\bf R}_j={\bf R}_i+{\bf a}_{x+y} \\
  0                       & \textrm{otherwise}
\end{array}
\right.,
\end{equation}
\end{itemize}
together with $a_0^1=a_1$, $a_0^2=a_2$, and $a_0^3=0$. For $\lambda=0$, this state becomes the $s$-wave state used in the 
square lattice.~\cite{gros1988} In contrast, for $\chi=\eta=0$ it corresponds to decoupled chains with $J=0$. It has been 
used to study the $J/J^\prime \ll 1$ limit in Ref.~\onlinecite{yunoki2006}, where it has gapless Dirac points.
 
In Eqs.~(\ref{eq:SL1}-\ref{eq:SL2_bis}), $\lambda$, $\chi$, and $\eta$, as well as $a_0^{1,2,3}$ represent variational 
parameters to be optimized. The following $Z_2B$ spin liquid with a $2\times 1$ unit cell (denoted by 
$Z2B\tau^1 \tau^2_{-} \tau^3_{+}$ in Ref.~\onlinecite{zhou2003}):~\cite{notaZ2B}
\begin{equation}\label{eq:SL3}
U_{ij}=\left\{
\begin{array}{ll}
  \eta\tau^1+\chi\tau^3             & {\rm for} \;\;\; {\bf R}_j={\bf R}_i+{\bf a}_{x} \\
  (-1)^{i_x}(\eta\tau^1-\chi\tau^3) & {\rm for} \;\;\; {\bf R}_j={\bf R}_i+{\bf a}_{y} \\
  (-1)^{i_x}\lambda\tau^3           & {\rm for} \;\;\; {\bf R}_j={\bf R}_i+{\bf a}_{x+y} \\
  0                       & \textrm{otherwise}
\end{array}
\right.,
\end{equation}
with $a_0^{1,2,3}=0$, was found to improve the $U(1)$ Dirac state of Eq.~(\ref{eq:SL1}) on small lattice sizes, but its energy 
gain goes to zero as the size of the cluster increases, since the additional parameter $\eta$ becomes negligible. The other 
four $Z_2$ spin liquids classified in Ref.~\onlinecite{zhou2003} are not competing in energies with the previous ones. 
We also mention that the $U(1)$ uniform spin liquid defined by (denoted by $U1A\tau^0 \tau^0_{+} \tau^1_{+}$ 
in Ref.~\onlinecite{zhou2003}):
\begin{equation}\label{eq:SL4}
U_{ij}=\left\{
\begin{array}{ll}
  \chi\tau^3    & {\rm for} \;\;\; {\bf R}_j={\bf R}_i+{\bf a}_{x} \\
  \chi\tau^3    & {\rm for} \;\;\; {\bf R}_j={\bf R}_i+{\bf a}_{y} \\
  \lambda\tau^3 & {\rm for} \;\;\; {\bf R}_j={\bf R}_i+{\bf a}_{x+y} \\
  0                       & \textrm{otherwise}
\end{array}
\right.,
\end{equation}
with $a_0^{1,2}=0$ and $a_0^3=a_3$, does not give competing energies with the other ones in the whole phase diagram.

Finally, we have also included a short-range spin-spin Jastrow factor on top of the spin-liquid mean-field state; this allows
a significant energy gain, but does not induce a fictitious magnetic order, since the spin-spin correlations remain short 
ranged. We also mention that the energy of the $Z_2A$ spin liquids has been slightly improved by extending the range of the 
$U_{ij}$ up to the sixth neighbors.

\section{Results}\label{sec:results}

In this section, we present the results of our calculations, by separately considering the two regimes $J^\prime/J\le 1$ and 
$J/J^\prime\le 1$. In both cases, we first investigate different spin-liquid and magnetic {\it Ans\"atze} and then compare 
the best spin liquid with the best magnetically ordered wave function, in order to determine the nature of the ground state. 
We remark that our approach allows us to consider on the same level spin-liquid and magnetic wave functions, which are 
relevant when considering the anisotropic triangular lattice. 

\subsection{The $J^\prime/J\le 1$ case}

\begin{figure}
\includegraphics[width=1.0\columnwidth]{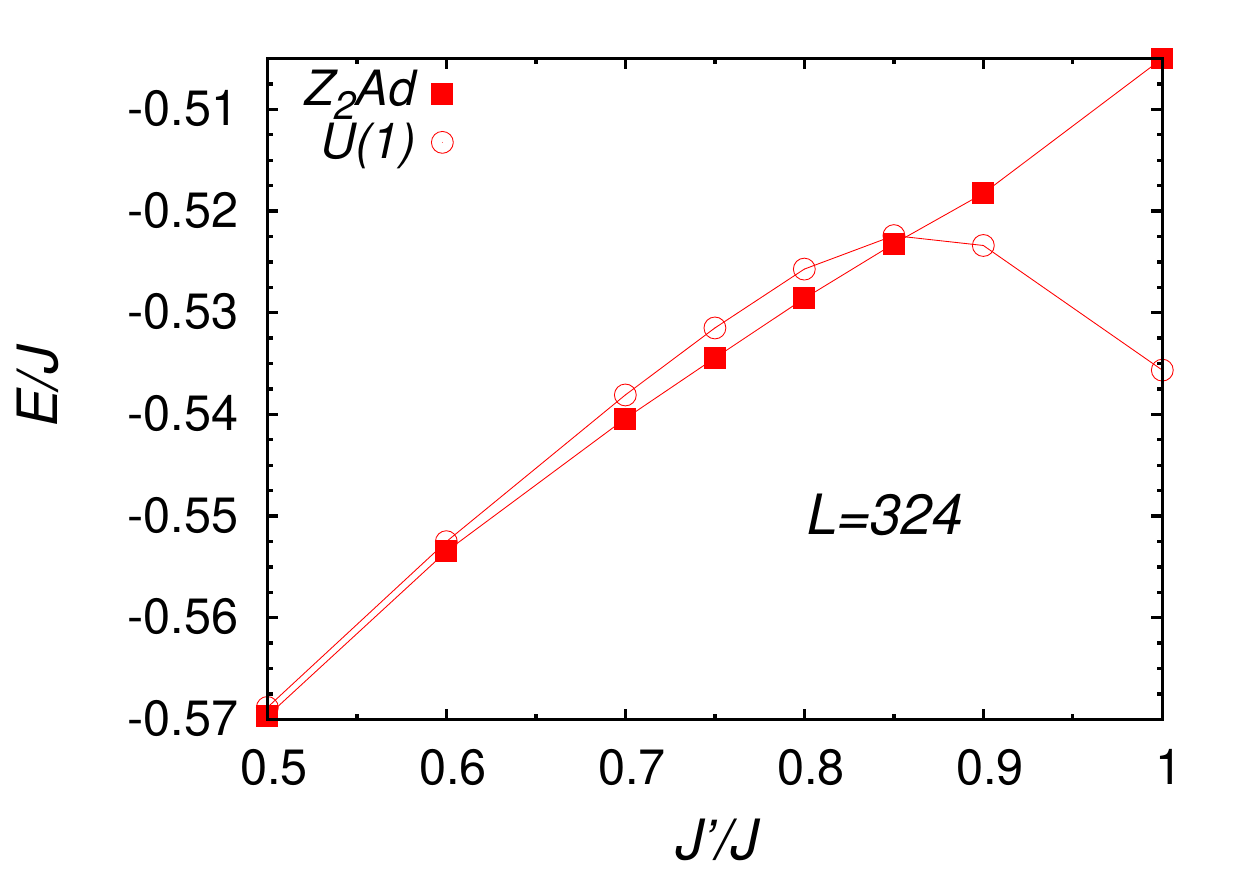}
\caption{\label{fig:SL_J>J'_18x18}
(Color online) Energy per site as a function of $J^\prime/J$ for two different spin liquids: the $Z_2Ad$ state of 
Eq.~(\ref{eq:SL2}) (red squares) and the $U(1)$ Dirac state of Eq.~(\ref{eq:SL1}) (red empty circles). All data are presented
on the $18 \times 18$ cluster.} 
\end{figure}

Let us start with spin-liquid states and consider a small $6 \times 6$ cluster. Here, we have systematically considered all 
the $U(1)$ and $Z_2$ spin liquids that have been classified in Ref.~\onlinecite{zhou2003}. On this small cluster, the energy 
of the $Z_2B$ spin liquid is slightly lower than the one of the $U(1)$ Dirac state, the energy gain being about 
$\Delta E/J \simeq 10^{-3}$ for $J^\prime/J=1$ and $\Delta E/J \simeq 10^{-4}$ for $J^\prime/J=0.7$. However, by increasing 
the size of the lattice, the energy gain strongly decreases and becomes negligible in the thermodynamic limit, thus indicating 
that the $U(1)$ Dirac state is stable. Nevertheless, we observe that the $Z_2Ad$ spin liquid of Eq.~(\ref{eq:SL2}) is favored 
over a broad range of frustrating ratios $J^\prime/J$, namely up to $J^\prime/J \simeq 0.9$. We would like to remark that no 
dimerization occurs in these states. In fact, there is no energy gain by allowing for hopping or pairing terms that break the 
translational symmetry. At difference with our results, a dimer-order state has been suggested to occur in the region 
$0.7\lesssim J^\prime/J\lesssim 0.9$.~\cite{weihong1999} However, the results of Ref.~\onlinecite{weihong1999} for 
non-magnetic phases are biased toward dimer order, since they are obtained by a series expansion calculation around 
dimerized states. The comparison between the energies of the $Z_2Ad$ and the $U(1)$ Dirac states for the $18 \times 18$ 
cluster is reported in Fig.~\ref{fig:SL_J>J'_18x18}. Also for this large cluster, the energy of the $Z_2Ad$ {\it Ansatz} 
is lower than the one of the $U(1)$ Dirac state for $J^\prime/J \lesssim 0.85$.

\begin{figure}
\includegraphics[width=1.0\columnwidth]{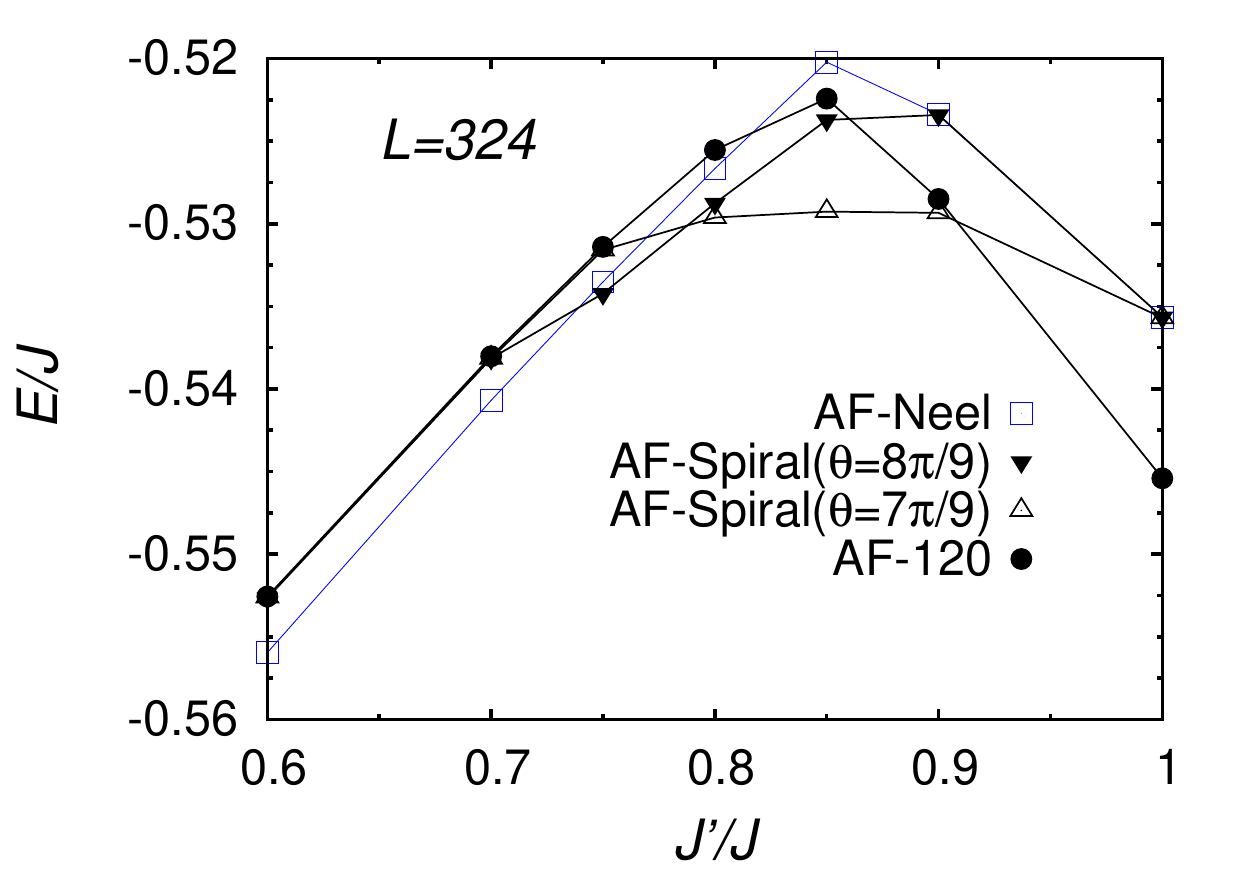}
\caption{\label{fig:AF_J>J'_18x18}
(Color online) Energy per site as a function of $J^\prime/J$ for four different magnetic wave functions: 
N\'eel order (blue empty squares), spiral magnetic order with $\theta=8\pi/9$ (black down triangles), spiral magnetic 
order with $\theta=7\pi/9$ (black empty up triangles), and $120^{\circ}$ order (black circles). 
All data are presented on the $18 \times 18$ cluster.} 
\end{figure}

Let us now move to magnetic wave functions. On the $6 \times 6$ lattice, only two states are relevant for $J^\prime/J\le 1$: 
the one with N\'eel order, where neighboring spins on the bonds with coupling $J$ form an angle $\theta=\pi$ (i.e., the
one obtained in the unfrustrated case with $J^\prime=0$), and the one with $120^{\circ}$ order, where they form an angle of 
$\theta=2\pi/3$ (i.e., the one obtained for the isotropic limit $J^\prime/J=1$). While the N\'eel order is favored for 
$J^\prime/J \lesssim 0.85$, the state with the $120^{\circ}$ order has a lower energy close to the isotropic point, namely for 
$0.85 \lesssim J^\prime/J \le 1$. It should be emphasized that on this small cluster only few ${\bf Q}$ vectors are allowed
imposing periodic boundary conditions and, therefore, it is impossible to assess the stability of generic spiral states. 
A less trivial result concerns the non-translational invariant nature of the hopping, with $\pi$-fluxes threading up triangles
(as discussed in Sec.~\ref{sec:model}). Indeed, we observed that the N\'eel state with a translationally invariant hopping has 
always an energy higher than the N\'eel state with $\pi$-fluxes in the kinetic energy, even for the unfrustrated case.
For example, for $J^\prime=0$, the translationally invariant kinetic part of Eq.~(\ref{eq:hopping_1x1}) gives an energy per 
site $E/J=-0.67142(1)$, while implementing the $2 \times 1$ unit cell of Eq.~(\ref{eq:hopping_2x1}), we obtain 
$E/J=-0.67529(1)$.

The situation becomes more interesting when considering a larger $18 \times 18$ cluster, see Fig.~\ref{fig:AF_J>J'_18x18}. 
Here, two intermediate magnetic orderings (interpolating the N\'eel and the $120^{\circ}$ orders) can be taken into account, 
with $\theta=8\pi/9$ and $\theta=7\pi/9$. Our results indicate that N\'eel order is favored up to $J^\prime/J \simeq 0.7$, 
intermediate spiral orders appear for $0.75 \lesssim J^\prime/J \lesssim 0.9$, while the $120^{\circ}$ order remains the 
magnetic state with the lowest energy close to the isotropic point (i.e., for $0.9 \lesssim J^\prime/J \le 1$). All these 
magnetic states are constructed with $\pi$-fluxes in the kinetic energy. Remarkably, the optimal antiferromagnetic field $h$ 
becomes large in the region where the corresponding magnetic order is energetically favored (not shown). Unfortunately, a 
precise determination of how the pitch angle $\theta$ changes with the frustration ratio $J^\prime/J$ is impossible, since 
it would require very large cluster sizes. Nevertheless, it is important to emphasize that we get a clear evidence of 
non-trivial magnetic orders (i.e., with angles $\theta \ne \pi$ or $2\pi/3$) already on this cluster.

\begin{figure}
\includegraphics[width=1.0\columnwidth]{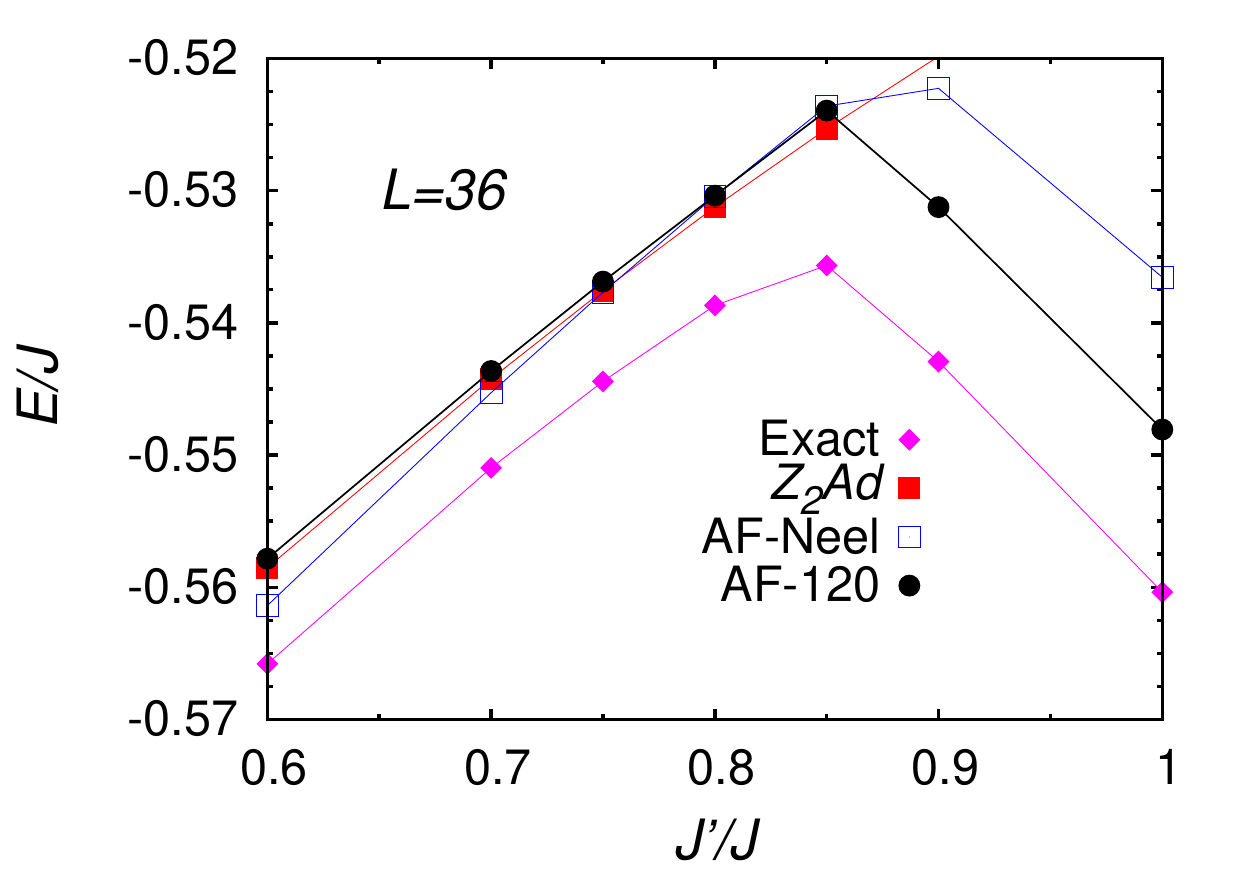}
\caption{\label{fig:PD_J>J'_6x6}
(Color online) Energy per site as a function of $J^\prime/J$ for three different wave functions: the $Z_2Ad$ spin liquid of 
Eq.~(\ref{eq:SL2}) (red squares), the magnetic states with N\'eel order (blue empty squares) and $120^{\circ}$ order (black 
circles). The exact results obtained by Lanczos diagonalization (magenta diamonds) are also reported for comparison.
All data are presented on the $6 \times 6$ cluster.}
\end{figure}

\begin{figure}
\includegraphics[width=1.0\columnwidth]{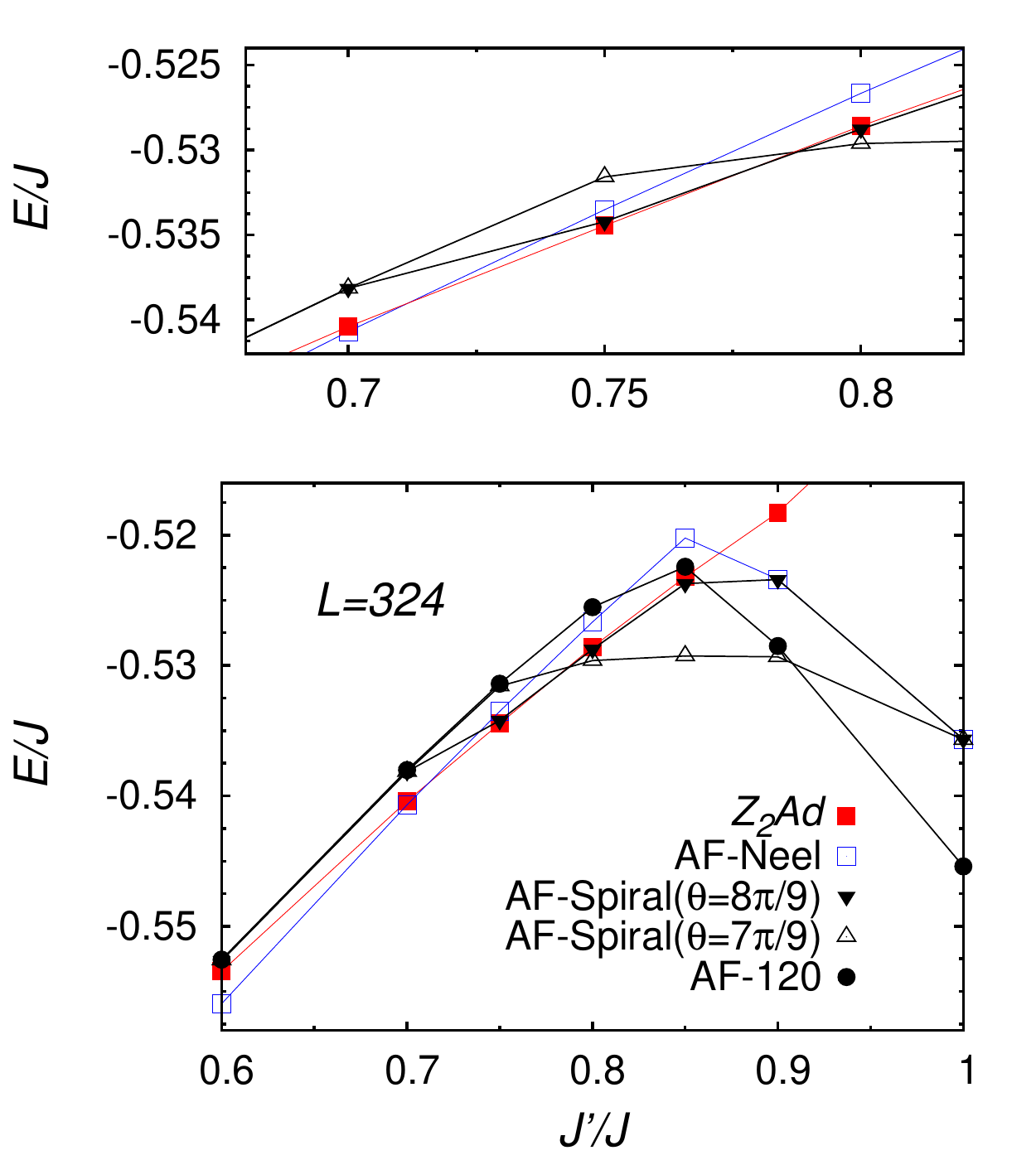}
\caption{\label{fig:PD_J>J'_18x18}
(Color online) Lower panel: Energy per site as a function of $J^\prime/J$ for five different wave functions: the $Z_2Ad$ 
spin liquid of Eq.~(\ref{eq:SL2}) (red squares), the magnetic states with N\'eel order (blue empty squares), spiral order 
with $\theta=8\pi/9$ (black down triangles) and $\theta=7\pi/9$ (black empty up triangles), and $120^{\circ}$ order (black  
circles). All data are presented on the $18 \times 18$ cluster. Upper panel: zoom of the highly-frustrated region 
$0.7 \lesssim J^\prime/J \lesssim 0.8$.} 
\end{figure}

Finally, we compare magnetic and spin-liquid phases. First of all, we show the results obtained on the $6 \times 6$ 
cluster, see Fig.~\ref{fig:PD_J>J'_6x6}. In this small lattice, exact results are available by the Lanczos technique,
thus providing the overall accuracy of these variational states. We notice that the exact ground-state energy reaches its 
maximum for $J^\prime/J \simeq 0.85$, which should mark the strongest possible frustration in this parameter regime. 
Here, being spiral states not available on this cluster, we can identify three different regions: two regimes in which 
magnetic wave functions prevail over spin-liquid ones, for $J^\prime/J \lesssim 0.75$ (N\'eel order) and 
$0.85 \lesssim J^\prime/J \le 1$ ($120^{\circ}$ order), and an intermediate region where the $Z_2Ad$ spin-liquid wave function
has the lowest variational energy, for $0.75 \lesssim J^\prime/J \lesssim 0.85$. However, it is precisely in this regime
that our wave functions have the worst accuracy. Indeed, the situation is quite different on larger system sizes, where
spiral states are also available. The main result is reported in Fig.~\ref{fig:PD_J>J'_18x18}, where the various variational
energies are reported for the $18 \times 18$ cluster. For the same cluster, we additionally report in Table~\ref{tab:energyI} 
the energies of the best spin-liquid and of the best magnetic state for $0.5 \le J^\prime/J \le 1$. Here, we can identify 
four regions of the phase diagram: the most interesting one appears for $0.7 \lesssim J^\prime/J \lesssim 0.8$, where the 
$Z_2Ad$ spin liquid is now challenged by the spiral magnetic order with $\theta=8\pi/9$, see also the upper panel of 
Fig.~\ref{fig:PD_J>J'_18x18}. We would like to mention that these values of the frustrating ratio correspond approximately to 
the ones where a spin-liquid region has been identified by a previous VMC study of the Hubbard model on the same lattice 
geometry,~\cite{tocchio2013} possibly suggesting that charge fluctuations can favor the spin-liquid state over the magnetic 
one. In contrast, the ground state can be clearly determined in the remaining regions of the phase diagram and is characterized
by N\'eel magnetism for $J^\prime/J \lesssim 0.7$, spiral order with $\theta=7\pi/9$ for $0.8 \lesssim J^\prime/J \lesssim 0.9$
and $120^{\circ}$ order for $0.9 \lesssim J^\prime/J \le 1$. We emphasize that these results have been obtained on a finite 
cluster, where only few pitch vectors are allowed. In the thermodynamic limit, we expect that the pitch angle of the magnetic 
phase varies continuously from $\pi$ to $2\pi/3$. In this regard, two remarks are appropriate: from the one side, we expect 
that the N\'eel phase with collinear order is more robust than in the classical limit, where it is stable for $J^\prime/J<1/2$.
Indeed, we expect that spiral phases with generic incommensurate order are much more fragile than collinear phases, with the 
value $J^\prime/J \simeq 0.7$ being a reasonable estimation for placing the transition from the N\'eel to the spiral phases. 
On the other side, it is extremely hard to clarify whether the $120^\circ$ order can remain stable away from the isotropic 
point or not. In fact, even though this simple coplanar state should have a larger stiffness with respect to generic 
(incommensurate) spirals, the determination of the precise periodicity close to $J^\prime/J=1$ would require huge clusters 
and massive numerical computations.

\begin{table}
\caption{\label{tab:energyI}
Energies for our best spin liquid and for our best magnetic state in the range $0.5 \le J^\prime/J \le 1$, on the
$18\times 18$ cluster.}
\begin{tabular}{ccc}
\hline
$J^\prime/J$ & $E/J$ (Spin liquid) & $E/J$ (Magnetic state) \\
\hline\hline
0.5          & -0.56967(1)         & -0.57233(1) \\
0.6          & -0.55347(1)         & -0.55595(1) \\
0.7          & -0.54040(1)         & -0.54069(1) \\
0.75         & -0.53445(1)         & -0.53422(1) \\
0.8          & -0.52859(1)         & -0.52963(2) \\
0.85         & -0.52322(1)         & -0.52927(1) \\
0.9          & -0.52339(2)         & -0.52934(1) \\
1.0          & -0.53565(2)         & -0.54542(1) \\
\hline \hline 
\end{tabular}
\end{table}

\subsection{The $J/J^\prime\le 1$ case}

\begin{figure}
\includegraphics[width=1.0\columnwidth]{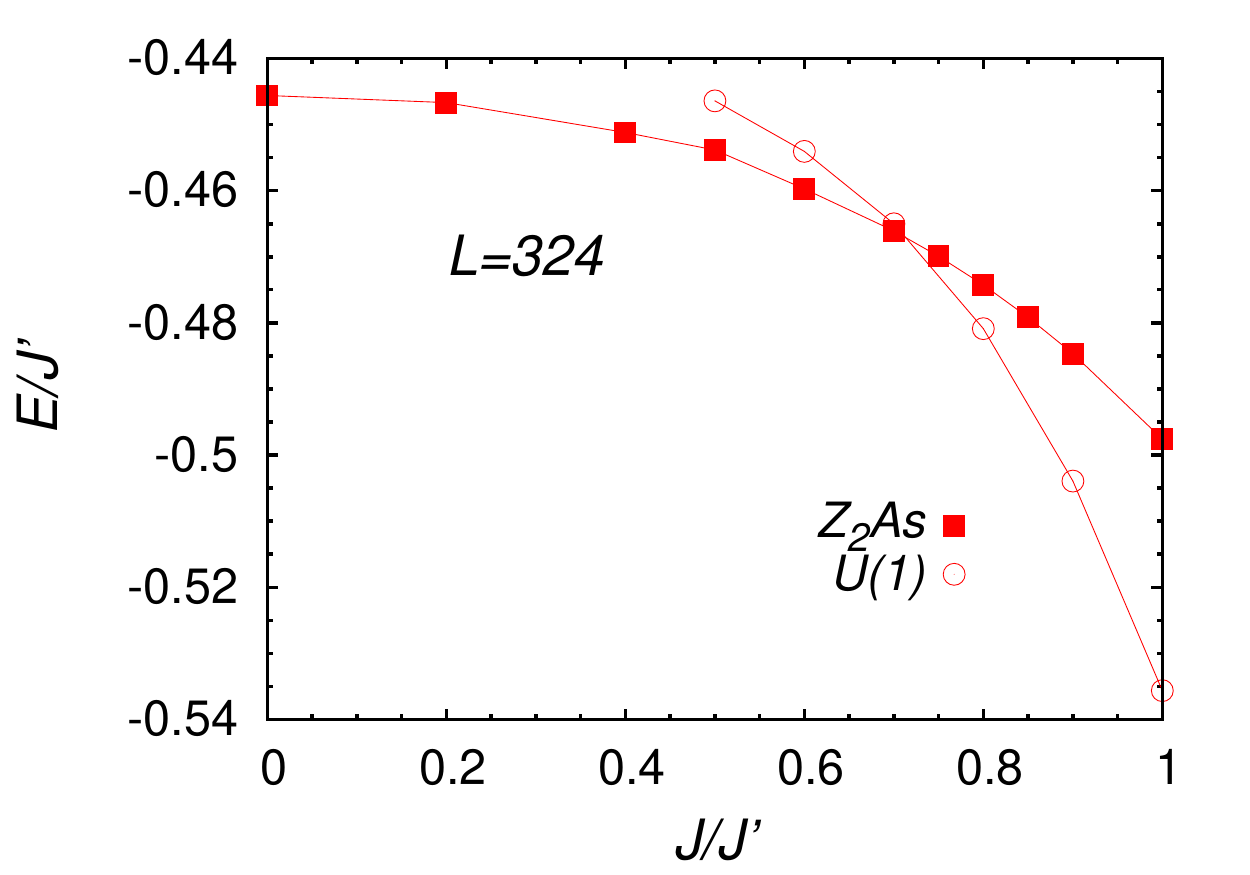}
\caption{\label{fig:SL_J'>J_18x18}
(Color online) Energy per site as a function of $J/J^\prime$ for two spin liquids: the $Z_2As$ state of Eq.~(\ref{eq:SL2_bis}) 
(red squares) and the $U(1)$ Dirac state of Eq.~(\ref{eq:SL1}) (red empty circles). All data are presented on the 
$18 \times 18$ cluster.} 
\end{figure}

In analogy to what we presented in the previous subsection, we start to investigate the $J/J^\prime\le 1$ case by considering 
the spin-liquid wave functions. The energies per site for the $18 \times 18$ cluster are shown in Fig.~\ref{fig:SL_J'>J_18x18},
where we report the results for the $Z_2As$ spin liquid of Eq.~(\ref{eq:SL2_bis}), which gives the optimal {\it Ansatz} for 
$J/J^\prime \lesssim 0.7$, and the $U(1)$ Dirac spin liquid of Eq.~(\ref{eq:SL1}), which represents the optimal state close 
to the isotropic point (i.e., $0.7 \lesssim J/J^\prime \le 1$). As before, also in this case there is a small energy gain 
(that is maximal at the isotropic point) by lowering the symmetry of the $U(1)$ Dirac spin liquid to $Z_2B$ on the small 
$6\times 6$ cluster, while the energy gain becomes negligible by increasing the lattice size.

Then, we move to consider magnetic states. On the $6 \times 6$ cluster, only two relevant magnetic orderings may be realized:
 the $120^{\circ}$ order, which appears to be stable for $0.8 \lesssim J/J^\prime \le 1$, and the collinear one with 
$\theta^\prime=\pi$ and $\theta=0$ or $\pi$, which is found for $J/J^\prime \lesssim 0.8$. This change in magnetic order is 
accompanied by a change in the effective dimensionality of the system: while the $120^{\circ}$ state is a true two-dimensional 
order, in the collinear one the only relevant interactions are the antiferromagnetic ones along the chains with coupling 
$J^\prime$, while along the bonds with coupling $J$ the spins align alternatively in a ferro- or antiferromagnetic way, see 
Fig.~\ref{fig:orders} right panel. The hopping structure of the magnetic state also changes from the one of 
Eq.~(\ref{eq:hopping_2x1}) for the $120^{\circ}$ order to the one of Eq.~(\ref{eq:hopping_1x1}) for the collinear one. 
In fact, the collinear order cannot coexist with $\pi$-fluxes in the kinetic energy, since this wave function turns out to 
have a negligible antiferromagnetic field $h$, with the only contribution to the variational energy being given by the kinetic
term. 

\begin{figure}
\includegraphics[width=1.0\columnwidth]{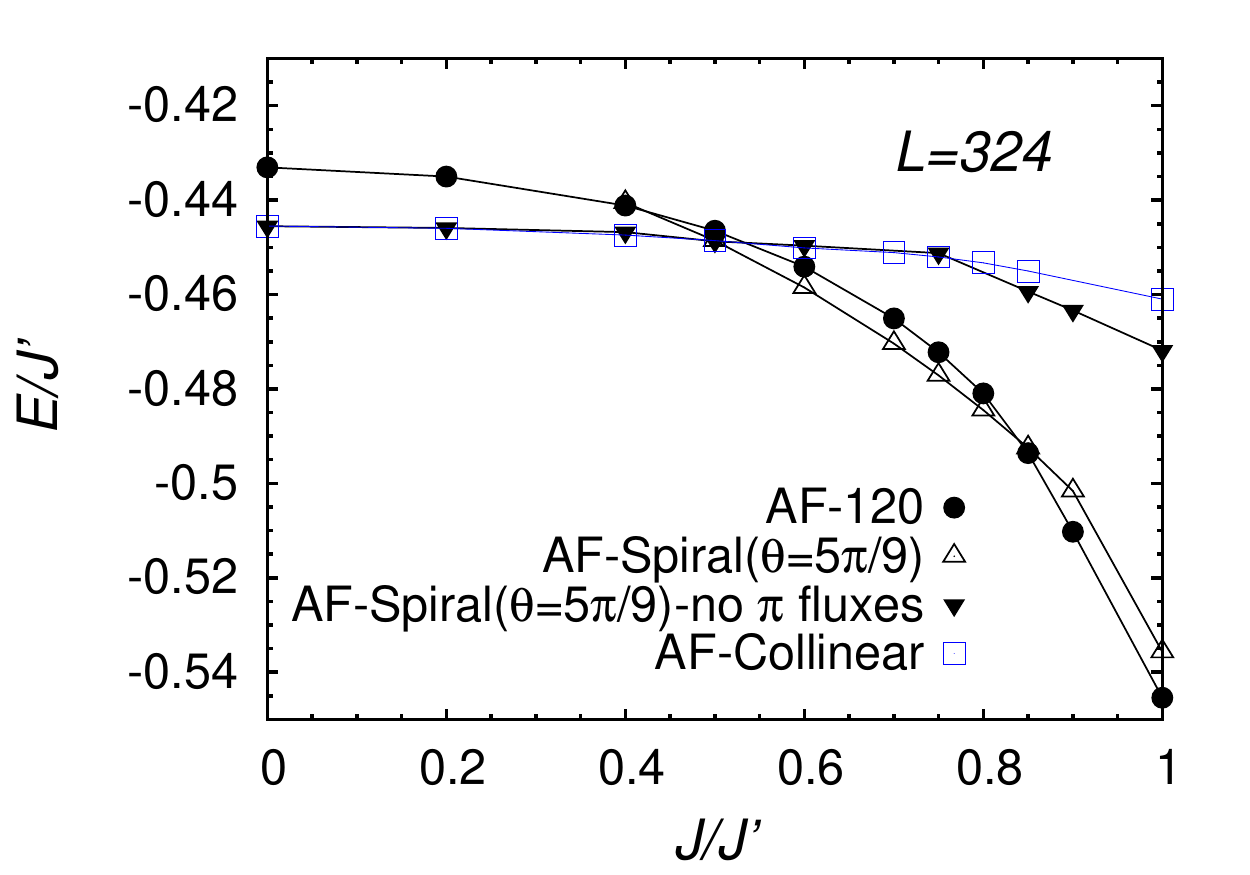}
\caption{\label{fig:AF_J'>J_18x18}
(Color online) Energy per site as a function of $J/J^\prime$ for four different magnetic wave functions: 
$120^{\circ}$ order with the hopping structure of Eq.~(\ref{eq:hopping_2x1}) (black circles), spiral order with 
$\theta=5\pi/9$ and the hopping structure of Eq.~(\ref{eq:hopping_2x1}) (black empty up triangles) and of 
Eq.~(\ref{eq:hopping_1x1}) (black down triangles), and collinear order with the hopping structure of Eq.~(\ref{eq:hopping_1x1})
(blue empty squares). All data are presented on the $18 \times 18$ cluster.} 
\end{figure}

On the $18 \times 18$ cluster, one further spiral order can be taken into account, i.e., the one with pitch angle 
$\theta=5\pi/9$. In Fig.~\ref{fig:AF_J'>J_18x18}, we show the variational energies for this state, in addition to the ones of
the states with $120^{\circ}$ and collinear order. Our results indicate that the spiral magnetic order, together with 
$\pi$-fluxes in the kinetic energy, is the best one in the range $0.5 \lesssim J/J^\prime \lesssim 0.85$. The importance of
having this non-trivial hopping structure close to the isotropic point is clear from the fact that the same magnetic order on 
top of the uniform hopping {\it Ansatz} of Eq.~(\ref{eq:hopping_1x1}) gives a much higher energy. Instead, for 
$J/J^\prime \lesssim 0.5$, the best spiral state is obtained with no magnetic fluxes piercing the lattice. Here, the energy 
of the spiral state with $\theta=5\pi/9$ is very close to the one obtained from collinear magnetism, indicating that the pitch
vector is not so relevant and, consequently, the ground state could be magnetically disordered.

\begin{figure}
\includegraphics[width=1.0\columnwidth]{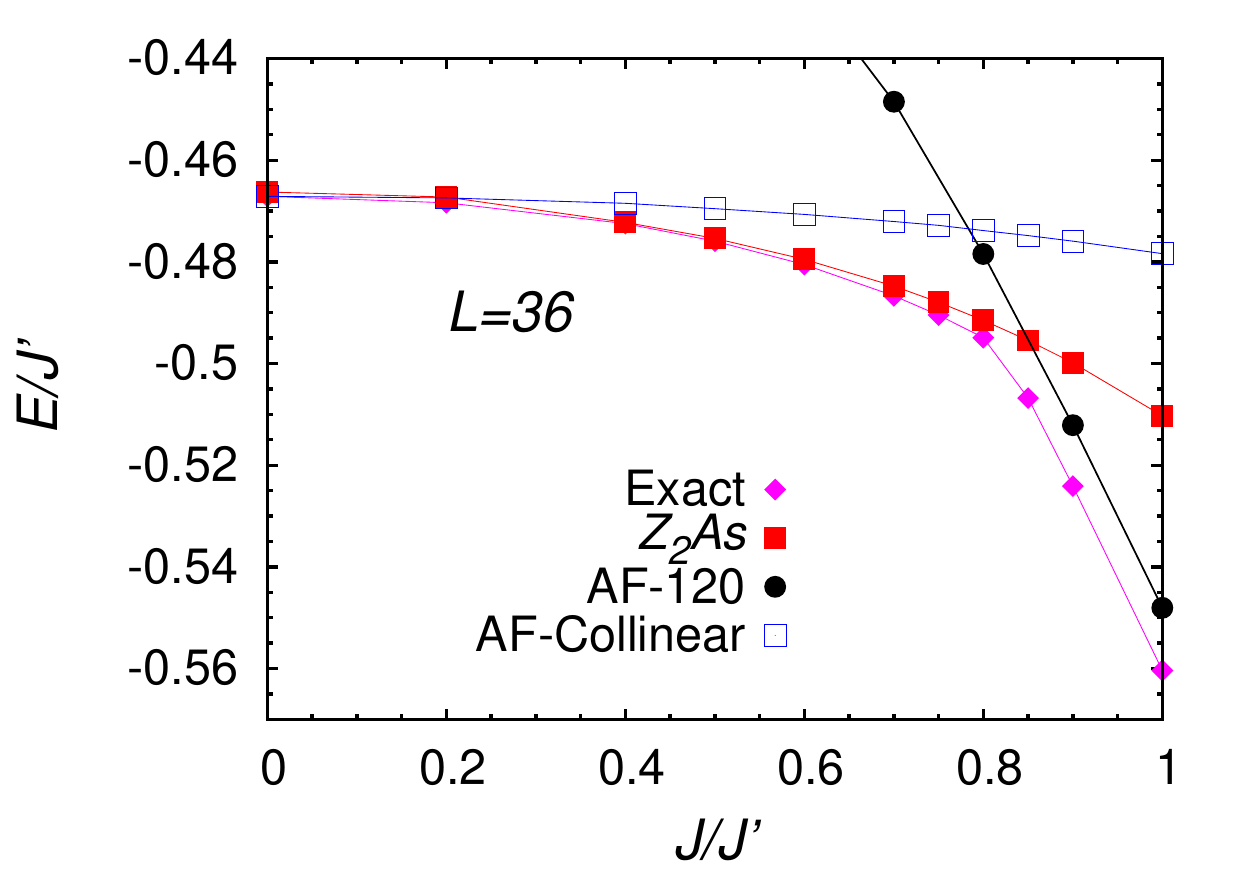}
\caption{\label{fig:PD_J'>J_6x6}
(Color online) Energy per site as a function of $J/J^\prime$, for three wave functions: the $Z_2As$ state of 
Eq.~(\ref{eq:SL2_bis}) (red squares), the $120^{\circ}$ ordered state with the hopping structure of Eq.~(\ref{eq:hopping_2x1}) 
(black circles), and collinear order with the hoppings of Eq.~(\ref{eq:hopping_1x1}) (blue empty squares). The exact 
results obtained by Lanczos diagonalization (magenta diamonds) are also reported for comparison. All data are presented on the 
$6 \times 6$ cluster.}
\end{figure}

\begin{figure}
\includegraphics[width=1.0\columnwidth]{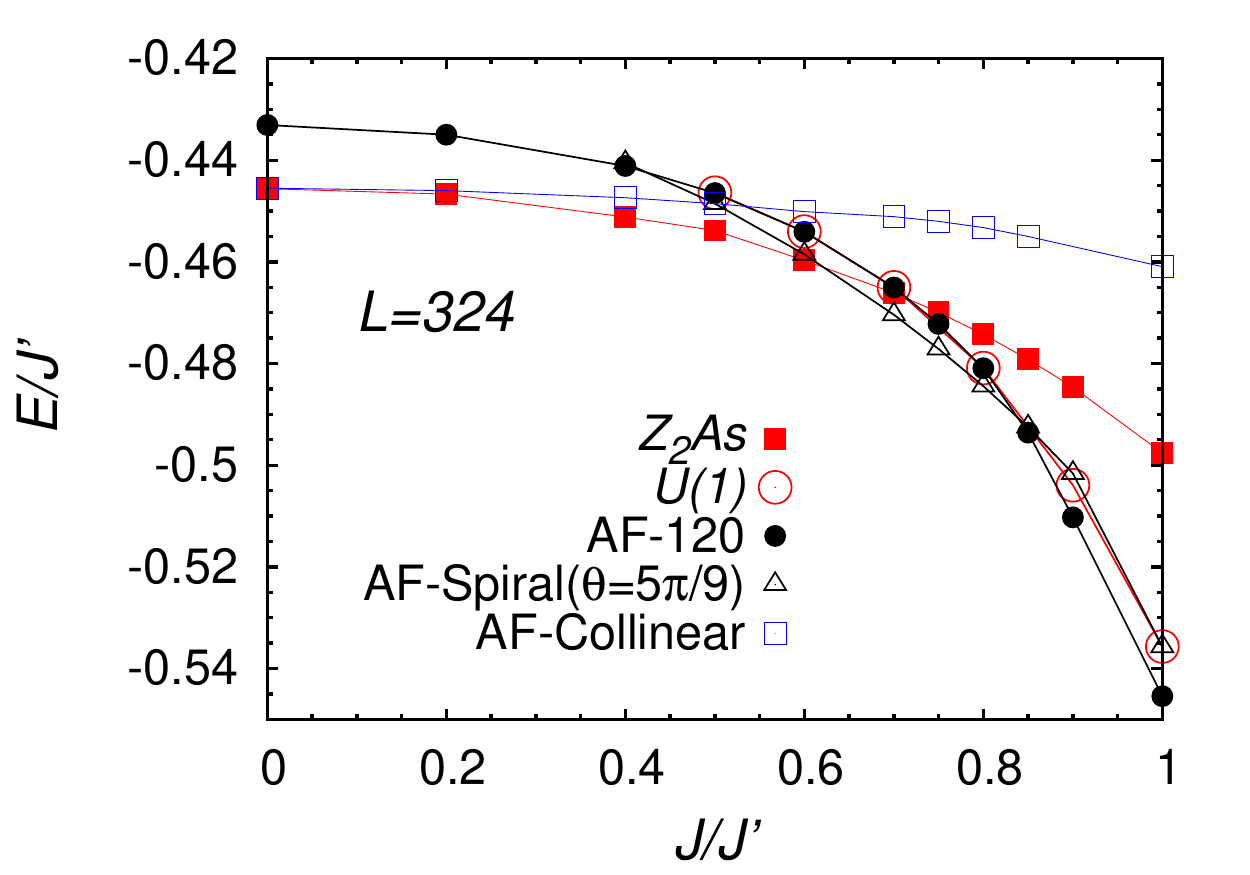}
\caption{\label{fig:PD_J'>J_18x18}
(Color online) Energy per site as a function of $J^\prime/J$ for five wave functions: the $Z_2As$ state of 
Eq.~(\ref{eq:SL2_bis}) (red squares), the $U(1)$ Dirac state of Eq.~(\ref{eq:SL1}) (red empty circles), the magnetic state 
with $120^{\circ}$ magnetism and the hoppings of Eq.~(\ref{eq:hopping_2x1}) (black circles), the one with spiral order 
with $\theta=5\pi/9$ and the hoppings of Eq.~(\ref{eq:hopping_2x1}) (black empty up triangles), and collinear order with 
the hoppings of Eq.~(\ref{eq:hopping_1x1}) (blue empty squares). All data are presented on the $18 \times 18$ lattice size.} 
\end{figure}

The comparison between magnetic and spin-liquid states on the $6 \times 6$ lattice is reported in Fig.~\ref{fig:PD_J'>J_6x6}, 
where exact results by the Lanczos technique are also shown. On this small cluster size, we can identify only two ground 
states: a magnetically ordered one for $0.85 \lesssim J/J^\prime \le 1$ and a spin-liquid state for $J/J^\prime \lesssim 0.85$;
in fact, in the $6 \times 6$ cluster, the collinear magnetic order does not give the lowest energy in any region of the
phase diagram. Notice that the spin-liquid wave function has a very good accuracy, compared with exact results, when 
frustration is not too large. The comparison between spin-liquid and magnetic states on the $18 \times 18$ cluster is finally
reported in Fig.~\ref{fig:PD_J'>J_18x18}. As a function of $J/J^\prime$, we identify three different ground states: the $Z_2As$
spin liquid of Eq.~(\ref{eq:SL2_bis}) for $J/J^\prime \lesssim 0.6$, the spiral order with $\theta=5\pi/9$ for 
$0.6 \lesssim J/J^\prime \lesssim 0.85$, and the $120^{\circ}$ magnetic order for $0.85 \lesssim J/J^\prime \le 1$. 
This result is different from previous variational ones,~\cite{yunoki2006,heidarian2009} especially close to the isotropic
point, where an additional gapped spin liquid (with a $2 \times 1$ unit cell) has been proposed for 
$0.65 \lesssim J/J^\prime \lesssim 0.8$. However, these previous calculations did not take into account magnetic spiral states,
which are expected to be relevant in this regime. We present in Table~\ref{tab:energyII} our best energies for the spin liquid 
and for the magnetic wave functions, compared with the optimal VMC energies of Refs.~\onlinecite{yunoki2006} 
and~\onlinecite{heidarian2009}. In both previous works, the ground state has been predicted to be a gapless spin liquid for 
$J/J^\prime \lesssim 0.65$, a gapped one for $0.65 \lesssim J/J^\prime \lesssim 0.85$, and a magnetic state for 
$0.85 \lesssim J/J^\prime \le 1$. Two remarks can be drawn from these data: (i) Our energies for the magnetic state are lower 
than the VMC optimal energies of Refs.~\onlinecite{yunoki2006,heidarian2009}, in the region where a gapped spin liquid state 
has been proposed, i.e., for $J/J^\prime=0.7$ and $0.8$; (ii) the optimal energies of Ref.~\onlinecite{heidarian2009} are 
rather accurate in the whole parameter range. This is due to the fact that they were obtained by using a full optimization of 
the pairing function and the Jastrow factor in real space. However, in this case, even if some evidence of incommensurate 
spiral order has been obtained, it was difficult to determine whether the wave function was really magnetically ordered or not.
The advantage of our present approach is given by the transparent representation of the variational wave functions, which 
describe either magnetic states or spin liquids.

Finally, we stress the fact that our present results confirm the existence of a quasi-one-dimensional spin liquid for
$J/J^\prime \lesssim 0.6$: although the collinear ordered state is quite close in energy, the $Z_2As$ (gapless) spin liquid
{\it Ansatz} gives a clear lower energy in this regime (while they become almost degenerate for $J \to 0$).

\begin{figure}
\includegraphics[width=0.9\columnwidth]{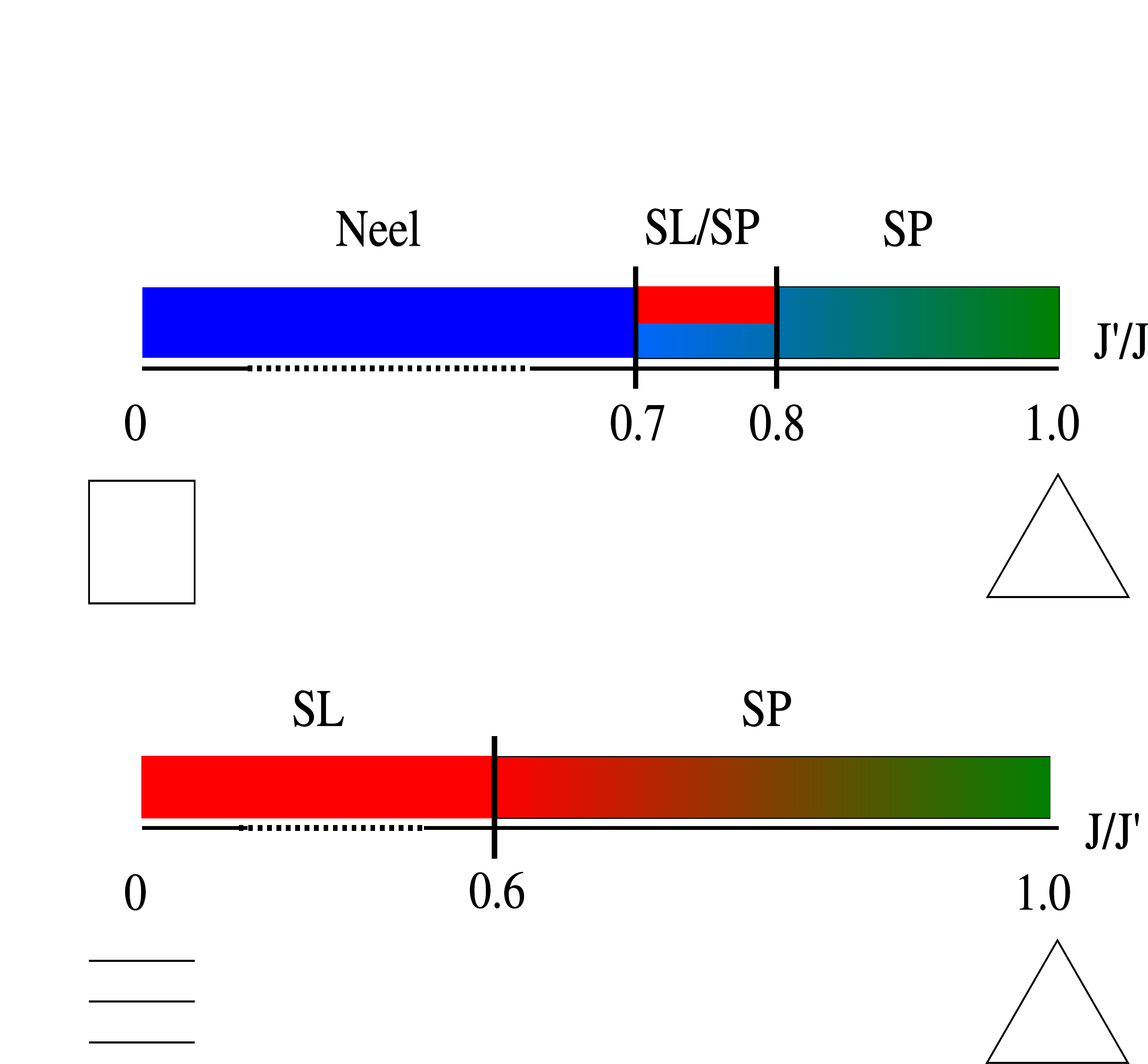}
\caption{\label{fig:pd}
(Color online) Our schematic VMC phase diagram of the Heisenberg model on the anisotropic triangular lattice. The regimes
where the gapless spin liquid and the spiral antiferromagnet phase are stabilized are denoted by SL and SP, respectively. 
The region where the N\'eel order is obtained is also reported. Finally, the region marked with both SL and SP 
indicates the place where these two phases are competing with very close energies.}
\end{figure}

\begin{table*}
\caption{\label{tab:energyII}
Energies for our best spin-liquid and for our best magnetic state in the range $0.2 \le J/J^\prime \le 1$, compared with 
the optimal energies of Ref.~\onlinecite{yunoki2006} and of Ref.~\onlinecite{heidarian2009}, on the $18\times 18$ cluster.}
\begin{tabular}{ccccc}
\hline
$J/J^\prime$ & $E/J^\prime$ (Spin liquid) & $E/J^\prime$ (Magnetic state) & $E/J^\prime$ (Ref.~\onlinecite{yunoki2006}) & $E/J^\prime$ (Ref.~\onlinecite{heidarian2009}) \\
\hline\hline
0.2 & -0.44668(1) & -0.44598(1) & -0.44687(1) & -0.44691(1) \\
0.4 & -0.45117(1) & -0.44740(1) & -0.45118(2) & -0.45127(1) \\
0.5 & -0.45384(1) & -0.44856(1) & -0.45474(2) & -0.45530(2) \\
0.6 & -0.45973(1) & -0.45850(1) & -0.45932(2) & -0.46048(2) \\
0.7 & -0.46610(1) & -0.47039(1) & -0.46514(2) & -0.46938(3) \\
0.8 & -0.48088(2) & -0.48445(1) & -0.47840(3) & -0.48369(3) \\
0.9 & -0.50394(2) & -0.51024(1) & -0.50007(3) & -0.51195(2) \\
1.0 & -0.53565(2) & -0.54542(1) & -0.53570(1) & -0.54716(3) \\
\hline\hline 
\end{tabular}
\end{table*}

\section{Conclusions}\label{sec:conc}

In this paper, we performed a systematic VMC study of the Heisenberg model on the anisotropic triangular lattice, for 
both $J^\prime/J \le 1$ and $J/J^\prime \le 1$, namely going from the unfrustrated square lattice ($J^\prime=0$) to 
the isotropic triangular lattice ($J^\prime=J$) and then from it to a set of decoupled chains ($J=0$). In particular, 
we constructed correlated wave functions for spiral magnetic orders and compared them with spin-liquid states obtained 
from the fermionic projective symmetry group classification of Ref.~\onlinecite{zhou2003}. Given the fact that these two 
families of variational states are written with the same language (Abrikosov fermions and spin-spin Jastrow factor), 
magnetic and non-magnetic states are treated on the same ground. The final sketch of the VMC phase diagram is shown in 
Fig.~\ref{fig:pd}. 

Starting from the unfrustrated case with $J^\prime=0$ and increasing the frustrating ratio up to $J^\prime/J \simeq 0.7$, 
the ground state exhibits N\'eel order. Notice that, since at the classical level the N\'eel state is stable up to 
$J^\prime/J=0.5$, we have a clear indication that quantum fluctuations favor the collinear magnetic order against coplanar 
spirals. Then, for $0.7 \lesssim J^\prime/J \lesssim 0.8$, magnetic states with generic pitch vectors (along the border of 
the Brillouin zone) and a $Z_2$ gapless spin liquid have very similar energies. On the $18 \times 18$ cluster that has been 
mainly used in our numerical simulations, the spiral state has $\theta=8\pi/9$ and we cannot resolve the competition between 
this state and the spin-liquid one, the difference between their energies being of the order of $10^{-4}$. Finally, for 
$0.8 \lesssim J^\prime/J \lesssim 1$, the ground state is expected to have magnetic order, with a pitch angle that 
continuously changes and reaches $\theta=2\pi/3$ for $J^\prime/J=1$, as suggested by the color gradient in Fig.~\ref{fig:pd}. 
On the $18 \times 18$ cluster, we can only consider a further spiral state with $\theta=7\pi/9$ (besides the $120^\circ$ 
order with $\theta=2\pi/3$), which gives the best variational energy for $0.8 \lesssim J^\prime/J \lesssim 0.9$. 

Although the Heisenberg model is not fully appropriate to describe organic charge-transfer salts, which are only moderately 
correlated, we observe that the parameter region where the spin-liquid {\it Ansatz} is competitive with magnetic states 
corresponds to the regime that is relevant for the spin-liquid compound $\kappa$-(ET)$_2$Cu$_2$(CN)$_3$ and for the 
Pd(dmit)$_2$ salts that do not order magnetically.~\cite{superexchange} Since, for the Heisenberg model, our present results 
suggest that the system is predominantly magnetic, charge fluctuations are expected to play an important role in stabilizing 
a non-magnetic ground state.

On the other side of the phase diagram, starting from the isotropic triangular lattice and reducing the inter-chain 
coupling $J$, we expect to have spiral order in the vicinity of the isotropic point. On the $18 \times 18$ lattice, 
we clearly see the stabilization of a magnetic state with $\theta=5\pi/9$ for $0.6 \lesssim J/J^\prime \lesssim 0.85$. 
Unfortunately, due to the finite size of the cluster, it is extremely difficult to follow in detail the evolution of 
the pitch vector as a function of $J/J^\prime$. Nevertheless, we expect that spiral phases are stabilized for 
$0.6 \lesssim J/J^\prime \le 1$. For $J/J^\prime \lesssim 0.6$, we obtain a clear indication that a gapless spin-liquid phase 
can be stabilized, in agreement with previous VMC calculations.~\cite{yunoki2006,hayashi2007,heidarian2009} In contrast, our 
results suggest that the gapped spin liquid that has been proposed to appear close to the isotropic 
point~\cite{yunoki2006,heidarian2009} is defeated by magnetically ordered states. In order to further check this outcome, 
we have also performed a Green's function Monte Carlo (GFMC) calculation,~\cite{trivedi1990} with the fixed-node 
approximation,~\cite{tenhaaf1995} at $J/J^\prime=0.7$ and $0.8$, by using both the best spin-liquid and the best magnetic 
state as trial wave functions. Our results show that the energy obtained by applying the GFMC method on the magnetic trial 
state is slightly lower (orders of $10^{-3}J^\prime$) than both the one obtained by using a non-magnetic trial state and the 
one reported in Ref.~\onlinecite{yunoki2006} (where the gapped spin-liquid state has been originally proposed). 
Therefore, our present results do not support the presence of a gapped spin liquid when $0.65 \lesssim J/J^\prime \le 1$.

Our finding that a gapless spin liquid is present in the weakly-coupled chain limit $J/J^\prime \lesssim 0.6$ is compatible 
with previous claims on Cs$_2$CuCl$_4$, which shows no magnetic order down to very small temperatures, with presumably gapless
spin excitations, and is characterized by $J/J^\prime \simeq 0.33$. Our results are also compatible with the recently 
discovered spin-liquid material $\kappa$-(ET)$_2$B(CN)$_4$, where a coupling ratio $J/J^\prime \simeq 0.5$ has been extracted 
from magnetic susceptibility measurements. Finally, we notice that, according to our numerical results, there is a striking 
difference between the ground-state properties with $J/J^\prime \simeq 0.4$, which should be magnetically disordered (or at 
most with a small antiferromagnetism with collinear order), and $J/J^\prime \simeq 0.75$, which should correspond to spiral 
magnetic order. Since Cs$_2$CuBr$_4$ is marked by having incommensurate spin correlations, the larger value of $J/J^\prime$ 
seems to be more appropriate for its low-energy description.


We thank S. Sorella for very useful discussions and S. Yunoki for providing some data for comparison. 
The authors acknowledge support from the Grant PRIN 2010 2010LLKJBX.


\begin{thebibliography}{99}
\bibitem{lacroix2011} See for example, \emph{Introduction to Frustrated Magnetism: Materials, Experiments, Theory}, 
   edited by C. Lacroix, P. Mendels, and F. Mila (Springer, 2011).
\bibitem{kanoda2011} K. Kanoda and R. Kato, Annu. Rev. Condens. Matter Phys. {\bf 2}, 167 (2011).
\bibitem{powell2011} B.J. Powell and R.H. McKenzie, Rep. Prog. Phys. {\bf 74}, 056501 (2011).
\bibitem{shimizu2003} Y. Shimizu, K. Miyagawa, K. Kanoda, M. Maesato, and G. Saito, \prl {\bf 91}, 107001 (2003).
\bibitem{manna2010} R.S. Manna, M. de Souza, A. Bruhl, J.A. Schlueter, and M. Lang, \prl {\bf 104}, 016403 (2010).
\bibitem{kandpal2009} H. C. Kandpal, I. Opahle, Y.-Z. Zhang, H. O. Jeschke, and R. Valenti, \prl {\bf 103}, 067004 (2009).
\bibitem{nakamura2009} K. Nakamura, Y. Yoshimoto, T. Kosugi, R. Arita, and M. Imada, J. Phys. Soc. Jpn. {\bf 78}, 
   083710 (2009).
\bibitem{scriven2012} E.P. Scriven and B. J. Powell, \prl {\bf 109}, 097206 (2012).
\bibitem{koretsune2014} T. Koretsune and C. Hotta, \prb {\bf 89}, 045102 (2014). 
\bibitem{yoshida2015} Y. Yoshida, H. Ito, M. Maesato, Y. Shimizu, H. Hayama, T. Hiramatsu, Y. Nakamura, H. Kishida, 
   T. Koretsune, C. Hotta, and G. Saito, Nat. Phys. {\bf 11}, 679 (2015).
\bibitem{jacko2013} A.C. Jacko, L.F. Tocchio, H.O. Jeschke, and R. Valent\'i, \prb {\bf 88}, 155139 (2013).
\bibitem{itou2008} T. Itou, A. Oyamada, S. Maegawa, M. Tamura, and R. Kato, \prb {\bf 77}, 104413 (2008).
\bibitem{tamura2006} M. Tamura, A. Nakao, and R. Kato, J. Phys. Soc. Jpn. {\bf 75}, 093701 (2006).
\bibitem{coldea2001} R. Coldea, D.A. Tennant, A.M. Tsvelik, and Z. Tylczynski, \prl {\bf 86}, 1335 (2001); 
   R. Coldea, D.A. Tennant, and Z. Tylczynski, \prb {\bf 68}, 134424 (2003).
\bibitem{vachon2011} M.-A. Vachon, G. Koutroulakis, V.F. Mitrovi\'c, O. Ma, J.B. Marston, 
A.P. Reyes, P. Kuhns, R. Coldea, and Z. Tylczynski, New J. Phys. {\bf 13}, 093029 (2011).
\bibitem{ono2004} T. Ono, H. Tanaka, O. Kolomiyets, H. Mitamura, T. Goto, K. Nakajima, A. Oosawa, Y. Koike, K. Kakurai, 
   J. Klenke, P. Smeibidle, and M. Meissner, J. Phys.: Condens. Matter {\bf 16}, S773 (2004).
\bibitem{zheng2005} W. Zheng, R.R.P. Singh, R.H. McKenzie, and R. Coldea, \prb {\bf 71}, 134422 
   (2005).
\bibitem{zvyagin2014}  S.A. Zvyagin, D. Kamenskyi, M. Ozerov, J. Wosnitza, M. Ikeda, T. Fujita, M. Hagiwara, 
   A.I. Smirnov, T.A. Soldatov, A.Ya. Shapiro, J. Krzystek, R. Hu, H. Ryu, C. Petrovic, and M.E. Zhitomirsky, 
   \prl {\bf 112}, 077206 (2014).
\bibitem{weihong1999} Z. Weihong, R.H. McKenzie, and R.R.P. Singh, \prb {\bf 59}, 14367 (1999).
\bibitem{ono2005} T. Ono, H. Tanaka, T. Nakagomi, O. Kolomiyets, H. Mitamura, F. Ishikawa, T. Goto, K. Nakajima, 
   A. Oosawa, Y. Koike, K. Kakurai, J. Klenke, P. Smeibidle, M. Meissner, and H. Aruga Katori, J. Phys. Soc. Jpn. 
   {\bf 74}, 135 (2005).
\bibitem{foyevtsova2011} K. Foyevtsova, I. Opahle, Y.-Z. Zhang, H.O. Jeschke, and
R. Valent\'i, Phys. Rev. B {\bf 83}, 125126 (2011).
\bibitem{bishop2009} R.F. Bishop, P.H.Y. Li, D.J.J. Farnell, and C.E. Campbell, \prb {\bf 79}, 174405 (2009).
\bibitem{powell2007} B.J. Powell and R.H. McKenzie, \prl {\bf 98}, 027005 (2007).
\bibitem{weichselbaum2011} A. Weichselbaum and S.R. White, \prb {\bf 84}, 245130 (2011).
\bibitem{bernu1994} B. Bernu, P. Lecheminant, C. Lhuillier, and L. Pierre, \prb {\bf 50}, 10048 (1994).
\bibitem{capriotti1999} L. Capriotti, A.E. Trumper, and S. Sorella, \prl {\bf 82}, 3899 (1999).
\bibitem{white2007} S.R. White and A.L. Chernyshev, \prl {\bf 99}, 127004 (2007).
\bibitem{merino1999} J. Merino, R.H. McKenzie, J.B. Marston, and C.H. Chung, J. Phys.: Condens. Matter {\bf 11}, 
   2965 (1999).
\bibitem{trumper1999} A.E. Trumper, \prb {\bf 60}, 2987 (1999).
\bibitem{hauke2011} P. Hauke, T. Roscilde, V. Murg, J.I. Cirac, and R. Schmied, New J. Phys. {\bf 13}, 075017 (2011).
\bibitem{holt2014} M. Holt, B.J. Powell, and J. Merino, \prb {\bf 89}, 174415 (2014).
\bibitem{manuel1999} L.O. Manuel and H.A. Ceccatto, \prb {\bf 60}, 9489 (1999).
\bibitem{yunoki2006} S. Yunoki and S. Sorella, \prb {\bf 74}, 014408 (2006).
\bibitem{hayashi2007} Y. Hayashi and M. Ogata, J. Phys. Soc. Jpn. {\bf 76}, 053705 (2007).
\bibitem{heidarian2009} D. Heidarian, S. Sorella, and F. Becca, \prb {\bf 80}, 012404 (2009).
\bibitem{weng2006} M.Q. Weng, D.N. Sheng, Z.Y. Weng, and R.J. Bursill, \prb {\bf 74}, 012407 (2006).
\bibitem{reuther2011} J. Reuther and R. Thomale, \prb {\bf 83}, 024402 (2011).
\bibitem{herfurth2013} T. Herfurth, S. Streib, and P. Kopietz, \prb {\bf 88}, 174404 (2013).
\bibitem{starykh2007} O.A. Starykh and L. Balents, \prl {\bf 98}, 077205 (2007).
\bibitem{chen2013} R. Chen, H. Ju, H.-C. Jiang, O.A. Starykh, and L. Balents, \prb {\bf 87}, 165123 (2013).
\bibitem{thesberg2014} M. Thesberg and E.S. S\o rensen, \prb {\bf 90}, 115117 (2014).
\bibitem{zhou2003} Y. Zhou and X.-G. Wen, arXiv:cond-mat/0210662.
\bibitem{lugas2006}  M. Lugas, L. Spanu, F. Becca, and S. Sorella, \prb {\bf 74}, 165122 (2006).
\bibitem{spanu2008} L. Spanu, M. Lugas, F. Becca, and S. Sorella, \prb {\bf 77}, 024510 (2008).
\bibitem{franjic1997} F. Franjic and S. Sorella, Prog. Theor. Phys. {\bf 97}, 399 (1997).
\bibitem{becca2000} F. Becca, M. Capone, and S. Sorella, \prb {\bf 62}, 12700 (2000).
\bibitem{ph} In the presence of pairing between up and down electrons, we perform a particle-hole 
   transformation in order to have a non-interacting Hamiltonian that commutes with the particle
   number, so as to define ``orbitals''.
\bibitem{gros1988} C. Gros, \prb {\bf 38}, 931(R) (1988).
\bibitem{zhang1988} F.C. Zhang, C. Gros, T.M. Rice, and H. Shiba, Supercond. Sci. Technol. {\bf 1}, 36 (1988).
\bibitem{notaZ2B} In order to highlight the connection with the $U(1)$ Dirac spin liquid of Eq.~(\ref{eq:SL1}), 
   we used a different gauge with respect to Ref.~\onlinecite{zhou2003}: $U_{ij} \to W_i^\dag U_{ij} W_j$ with
   $W_i=i/\sqrt{2}(\tau^2-\tau^3)$.
\bibitem{tocchio2013} L.F. Tocchio, H. Feldner, F. Becca, R. Valent\'i, and C. Gros, \prb {\bf 87}, 035143 (2013).
\bibitem{superexchange} Since the density-functional theory calculations estimate the hopping parameters $t$ and 
  $t^\prime$, we converted them into super-exchange couplings via the relations $J=4t^2/U$ and $J^\prime=4(t^\prime)^2/U$. 
\bibitem{trivedi1990} N. Trivedi and D.M. Ceperley, Phys. Rev. B {\bf 41}, 4552 (1990).
\bibitem{tenhaaf1995} D.F.B. ten Haaf, H.J.M. van Bemmel, J.M.J. van Leeuwen, W. van Saarloos, and D.M. Ceperley, 
 \prb {\bf 51}, 13039 (1995).
\end{thebibliography}
\end{document}